\pdfoutput=1  % force pdfLaTeX on arXiv (submission contains PDF figures)
\PassOptionsToPackage{table,dvipsnames,svgnames}{xcolor}
\documentclass[twocolumn]{fairmeta}

\usepackage{latexsym}
\usepackage[utf8]{inputenc}
\usepackage{amsmath}
\usepackage{amssymb}
\usepackage{pifont}
\usepackage{fontawesome}
\usepackage{listings}
\usepackage{colortbl}
\usepackage{array}
\usepackage{mdframed}
\usepackage{fancyvrb}
\usepackage{titletoc}
\usepackage[linesnumbered,ruled,vlined,algo2e]{algorithm2e}

\definecolor{skyblue}{RGB}{52, 152, 219}
\definecolor{lightskyblue}{RGB}{46, 204, 113}
\definecolor{sftgold}{RGB}{123, 104, 238}
\definecolor{genbg}{RGB}{225,245,225}
\definecolor{codekw}{RGB}{0,100,180}
\definecolor{systembg}{RGB}{220,235,255}
\definecolor{userbg}{RGB}{220,245,220}
\definecolor{assistbg}{RGB}{235,225,250}
\definecolor{skippedgray}{RGB}{240,240,240}
\definecolor{framecolor}{HTML}{B0B0B0}
\definecolor{kwblue}{HTML}{2563EB}
\definecolor{strred}{HTML}{B91C1C}
\definecolor{cmtgreen}{HTML}{16A34A}

\newcommand{\cmark}{{\color{green!90!black}\ding{51}}}
\newcommand{\xmark}{{\color{red!85!yellow}\ding{55}}}

\newtcblisting{systemprompt}[1]{%
  enhanced, breakable,
  colback=systembg, colframe=systembg!50!gray,
  boxrule=0.4pt, sharp corners,
  left=5pt, right=5pt, top=1pt, bottom=1pt, boxsep=2pt,
  before skip=2pt, after skip=2pt,
  before upper={\textbf{\small #1}\par\vspace{1pt}},
  listing only,
  listing options={basicstyle=\ttfamily\fontsize{7}{8.5}\selectfont,
                   breaklines=true, columns=fullflexible, keepspaces=true},
}

\newtcblisting{userprompt}[1]{%
  enhanced, breakable,
  colback=userbg, colframe=userbg!50!gray,
  boxrule=0.4pt, sharp corners,
  left=5pt, right=5pt, top=1pt, bottom=1pt, boxsep=2pt,
  before skip=2pt, after skip=2pt,
  before upper={\textbf{\small #1}\par\vspace{1pt}},
  listing only,
  listing options={basicstyle=\ttfamily\fontsize{7}{8.5}\selectfont,
                   breaklines=true, columns=fullflexible, keepspaces=true},
}

\newtcblisting{genprompt}[1]{%
  enhanced, breakable,
  colback=assistbg, colframe=assistbg!50!gray,
  boxrule=0.4pt, sharp corners,
  left=5pt, right=5pt, top=1pt, bottom=1pt, boxsep=2pt,
  before skip=2pt, after skip=2pt,
  before upper={\textbf{\small #1}\par\vspace{1pt}},
  listing only,
  listing options={basicstyle=\ttfamily\fontsize{7}{8.5}\selectfont,
                   breaklines=true, columns=fullflexible, keepspaces=true},
}

\lstdefinestyle{gdcode}{%
  basicstyle=\ttfamily\fontsize{7}{8.5}\selectfont,
  keywordstyle=\color{kwblue}\bfseries,
  stringstyle=\color{strred},
  commentstyle=\color{cmtgreen}\itshape,
  morekeywords={extends,func,var,const,signal,enum,if,else,elif,for,
                while,return,await,and,or,not,in,is,pass,true,false,
                null,self,onready,export},
  morecomment=[l]{\#},
  morestring=[b]",
  breaklines=true,
  columns=fullflexible,
  keepspaces=true,
  aboveskip=0pt,
  belowskip=0pt,
  xleftmargin=0pt,
  framexleftmargin=0pt,
  showstringspaces=false,
}

\newtcolorbox{turnbox}[1]{%
  enhanced, breakable,
  colback=#1, colframe=#1!50!gray,
  boxrule=0.4pt, sharp corners,
  left=5pt, right=5pt, top=1pt, bottom=1pt, boxsep=2pt,
  before skip=2pt, after skip=2pt,
}

\newtcolorbox{skipbox}{%
  enhanced, breakable,
  colback=skippedgray, colframe=skippedgray!50!gray,
  boxrule=0.4pt, sharp corners,
  left=5pt, right=5pt, top=4pt, bottom=4pt, boxsep=2pt,
  before skip=2pt, after skip=2pt,
}

\newtcolorbox{outerframe}[1][]{%
  enhanced, breakable,
  colback=white, colframe=framecolor,
  boxrule=0.5pt, arc=2pt,
  left=4pt, right=4pt, top=4pt, bottom=4pt, boxsep=0pt,
  before skip=4pt, after skip=4pt,
  #1
}

\newcommand{\turnlabel}[1]{\textbf{\small #1}\par\vspace{1pt}}

\title{\raisebox{-0.18\height}{\includegraphics[height=1.05em]{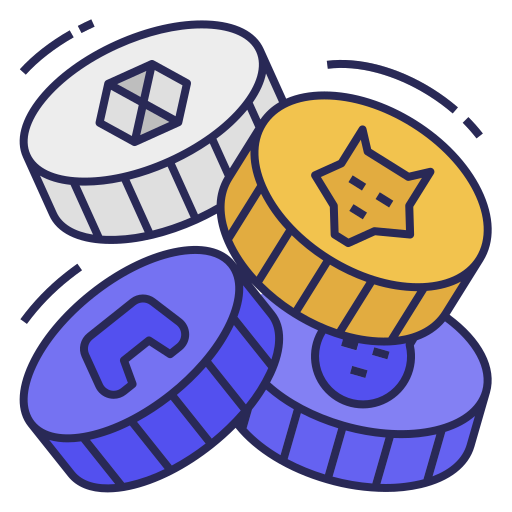}}\hspace{0.25em}\textit{J}AMER: Project-Level Code Framework Dataset and Benchmark on Professional Game Engines}

\author[1,2,3]{Jianwen Sun}
\author[1]{Chuanhao Li}
\author[1,2,3]{Zizhen Li}
\author[1,2,3]{Yukang Feng}
\author[3]{Fanrui Zhang}
\author[1]{Yifei Huang}
\author[2]{Yu Dai}
\author[{1,3}\,\protect\mbox{\footnotesize\faEnvelope}]{Kaipeng Zhang}

\affiliation[1]{Alaya Studio}
\affiliation[2]{Nankai University, Tianjin, China}
\affiliation[3]{Shanghai Innovation Institute, Shanghai, China}

\correspondence{\email{sunjianwen@mail.nankai.edu.cn}, \email{kaipeng.zhang@shanda.com}}
\date{\today}

\abstract{
Current AI-driven game development has made substantial progress in asset generation, gameplay design, and web-based game coding, yet project-level code engineering on professional game engines remains largely unexplored due to the absence of large-scale datasets and deterministic evaluation methods.
We present JamSet and JamBench, the first project-level game code framework dataset and benchmark built on a professional game engine.
Our key insight is that Game Jam competitions, community events where developers build complete games under tight time constraints, yield thousands of open-source projects suitable for this purpose.
Building on the Godot engine's text-based format and headless execution mode, we design a deterministic verification pipeline from file integrity to runtime behavior collection, distilling 8,133 verified projects from over 240,000 repositories.
Of these, 300 manually verified projects form JamBench; the rest constitute JamSet.
JamBench defines theme-driven generation and code completion tasks, evaluated through a pipeline combining compilation pass rates, Structural Completeness Score (SCS), and Behavioral Alignment Score (BAS).
Evaluation of 9 frontier models reveals a capability cliff as project scale increases, with runtime pass rates dropping from 80.4\% on small projects to 5.7\% on large ones (Task2a).
Code Agents improve compilation rates yet yield no gains in runtime behavioral quality, indicating that the bottleneck lies in architectural design rather than syntactic correctness. Experiments validate JamSet as effective training data. All data and code are publicly available.
}

\begin{document}
\maketitle

\begin{figure}[t]
\centering
\includegraphics[width=0.85\columnwidth]{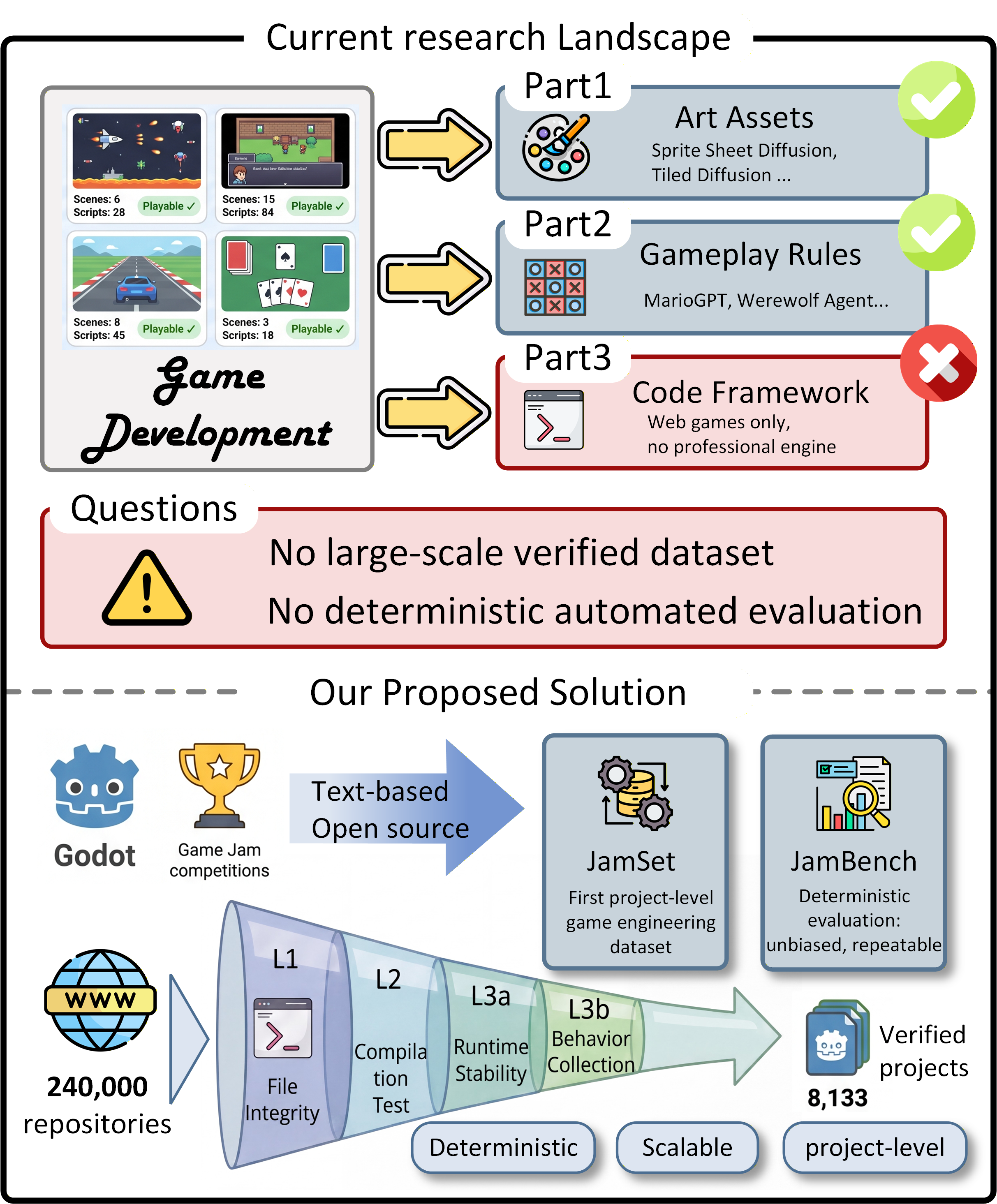}
\caption{Research landscape and our proposed solution. Upper:three layers, with code framework largely unexplored. Lower: leveraging Godot, we build JamSet and JamBench with a deterministic verification pipeline.}
\label{fig:teaser}
\end{figure}

\section{Introduction}
% Game development involves three layers: art assets, gameplay rules, and code frameworks. In recent years, AI has made significant progress on the first two: generative models produce high-quality textures and sprites~\cite{intro_1,intro_2,intro_3}, while procedural content generation (PCG)~\cite{intro_4,intro_5} and LLM-based rule design~\cite{intro_6,intro_7} continue to advance. In contrast, automated construction of code frameworks remains in its early stages: existing work primarily targets lightweight web frameworks generating 2D game code in JavaScript or TypeScript~\cite{intro_8,intro_9}, while the few efforts on professional game engines are limited to local modifications~\cite{intro_10}. \textbf{Project-level code framework generation on professional engines has received almost no systematic study}. Two factors contribute to this gap: the complexity of professional engine project structures~\cite{intro_11} poses a high barrier for LLM processing, and the runtime interactive nature of games makes traditional code evaluation methods difficult to apply~\cite{intro_12}, resulting in a lack of both large-scale datasets and reproducible evaluation methods (Figure~\ref{fig:teaser}).

Game development spans multiple interconnected concerns, from art assets and gameplay rules to underlying code frameworks. AI-driven approaches have made substantial progress on assets and gameplay: generative models produce high-quality textures and sprites\cite{intro_1,intro_2,intro_3}, while procedural content generation \cite{intro_4,intro_5} and LLM-based rule design \cite{intro_6,intro_7} continue to advance. On the code side, recent work has explored lightweight web frameworks that generate 2D games in JavaScript or TypeScript \cite{intro_8,intro_9}, and a few efforts target local modifications on professional engines \cite{intro_10}. However, \textbf{project-level code framework generation on professional game engines remains largely unexplored}, due to the absence of both large-scale datasets and deterministic evaluation methods.

Two interrelated challenges underlie this gap. The first is evaluation. In traditional software engineering, correctness can be verified through unit testing against expected outputs \cite{intro_13,intro_14}. However, game behavior is context-dependent at the project level, with no simple input-output correspondence \cite{intro_15}. Existing alternatives each fall short: hand-written test scripts are precise but prohibitively expensive to scale \cite{intro_10}; VLM- or LLM-based scoring offers scalability but introduces subjectivity and lacks reproducibility \cite{intro_8,intro_9}; and standard unit testing cannot capture game-specific runtime behaviors \cite{intro_16,intro_17}. The second challenge is data. Without a scalable evaluation method, there is no reliable way to filter high-quality projects from noisy open-source repositories, making large-scale dataset construction infeasible. Together, the absence of a deterministic evaluation framework makes it difficult to curate high-quality datasets or objectively assess model-generated games (Figure~\ref{fig:teaser}).

The Godot engine offers a unique opportunity to bridge this gap. Unlike engines that rely on binary formats and graphical interfaces \cite{intro_19}, Godot adopts a fully text-based project format: its scene files (.tscn), script files (.gd), and project configuration (.godot) are all human-readable plain text \cite{intro_10}, naturally compatible with LLM processing. Moreover, Godot is the fastest-growing open-source game engine, with steadily rising adoption in Game Jam competitions (Figure~\ref{fig:godot_growth}a). Game Jam competitions \cite{intro_20,intro_21,intro_22} provide an ideal data source: strict time constraints (48--72 hours) ensure entries are compact yet complete, and the open-source sharing tradition yields large-scale, freely accessible real-world data. Crucially, Godot's headless mode allows games to run without a graphical interface, enabling fully automated execution and runtime behavior collection at scale.

Building on these properties, we construct JamSet and JamBench. We first design a deterministic verification pipeline on Godot's headless mode, progressively checking file integrity, compilation correctness, runtime stability, and runtime behavior. This pipeline distills 8,133 verified projects from over 240,000 candidate repositories. Based on this, 300 are manually verified as the benchmark subset JamBench; the remaining 7,833 are processed into multi-turn training data as JamSet. JamBench covers theme-driven from-scratch generation (Task~1) and multi-granularity code completion (Task~2), evaluated through compilation pass rates, Structural Completeness Score (SCS) measuring static structural coverage, and Behavioral Alignment Score (BAS) measuring runtime behavioral similarity to real projects.

% Building on these findings, we construct JamSet and JamBench. Leveraging Godot's headless mode, we design a deterministic multi-level verification pipeline from code verification to runtime behavior collection. Based on this, we distill 8,133 verified game projects from over 240,000 repositories. From these, 300 projects are manually verified as the benchmark subset; the remaining 7,833 are processed into training data (JamSet). Fine-tuning experiments on JamSet validate the effectiveness of the dataset. Finally, we define two evaluation tasks, theme-driven from-scratch generation and multi-granularity code completion, with Structural Completeness Score (SCS) and Behavioral Alignment Score (BAS) as evaluation metrics (JamBench).

We evaluate 9 frontier models and 5 Code Agent configurations on JamBench. Results reveal a capability cliff as project scale increases \cite{intro_17,intro_26,intro_27}: on Task~2, runtime pass rates drop from 80.4\% on small projects to 5.7\% on large ones(Task2a). \textbf{Code Agents substantially improve compilation pass rates yet yield no gains in runtime behavioral quality} \cite{intro_28,intro_29}, indicating that the bottleneck lies in architectural design rather than syntactic correctness. Fine-tuning on JamSet validates the dataset's effectiveness: the base model shows improved compilation rates and structural completeness, while also adopting human engineering practices such as input abstraction and global state management.

% Our main contributions are: \textbf{(1)} A deterministic verification pipeline on a professional game engine, from code verification to runtime behavior collection. \textbf{(2) }The first project-level game code framework dataset JamSet (7,833 projects) and benchmark JamBench (300 projects). \textbf{(3)} Systematic evaluation revealing a sharp project-level capability cliff and the boundary of Code Agent between syntactic repair and semantic design.

Our main contributions are: \textbf{(1)} The first project-level game code framework dataset and benchmark on a professional game engine. \textbf{(2)} A deterministic verification pipeline from file integrity to runtime behavior collection. \textbf{(3)} Systematic evaluation revealing a capability cliff at project scale and the limitation of Code Agents to syntactic repair over architectural design.

\begin{table*}[t]
\centering
\small
\resizebox{\textwidth}{!}{%
\begin{tabular}{lccccccccc}
\toprule
\textbf{Work} & \textbf{Engine} & \textbf{Task} & \textbf{Scale} & \textbf{Evaluation} & \textbf{Determ.} & \textbf{Scalable} & \textbf{Proj.-level} & \textbf{Dataset} \\
\midrule
V-GameGym & Pygame & Single-file & 2,219 & LLM judge & \xmark & \cmark & \xmark & \cmark \\
OpenGame & Phaser/JS & From-scratch & 150 & VLM judge & \xmark & \cmark & \xmark & \xmark \\
GameDevBench & Godot & Local edit & 132 & Test scripts & \cmark & \xmark & \xmark & \xmark \\
AutoUE & UE5 & From-scratch & 20 & LLM judge & \xmark & \xmark & \cmark & \xmark \\
\midrule
\rowcolor{blue!4} Ours & Godot & Both & 8,133 & Engine verify & \cmark & \cmark & \cmark & \cmark \\
\bottomrule
\end{tabular}
}
\caption{Comparison of related works. ``Both'' refers to from-scratch generation and multi-granularity completion.}
\vspace{-2mm}
\label{tab:comparison}
\end{table*}

\section{Related Works}
% \textbf{游戏生成与评测 }AI 驱动的游戏创作发展迅速。GameTileNet 等工作聚焦游戏美术素材的生成，MarioGPT 等通过 LLM 实现关卡和玩法规则的设计。在代码框架层面：V-GameGym 等基于 Pygame 框架评测 LLM 的单文件游戏生成能力，使用 LLM 自动打分进行评估。OpenGame 构建了面向 Web 游戏（Phaser/JS）的 agentic 生成框架，训练了专用模型，并通过 VLM 打分进行评测。在专业游戏引擎方向，GameDevBench 定义了 132 个局部修改任务，通过手写测试脚本验证特定行为。AutoUE 在 Unreal Engine 5 上构建多 agent 系统实现端到端 3D 游戏生成，设计了自动化 play-testing 管道。UniGen 和 DreamGarden 分别在 Unity 和 Unreal Engine 上尝试了多 agent 代码生成和分层规划的游戏原型构建。这些工作推动了游戏代码框架生成从轻量框架走向专业引擎，但现有工作要么规模较小，要么评测依赖主观判断（VLM/LLM 打分），在大规模项目级代码框架生成与客观确定性评测方面仍存在空白。
\textbf{Game Generation and Evaluation.}
AI-driven game creation has advanced rapidly. GameTileNet~\cite{rw_1} and similar work focus on art asset generation, while MarioGPT~\cite{intro_6} leverages LLMs for gameplay design. On the code framework front, V-GameGym~\cite{intro_9} evaluates single-file game generation on Pygame, and OpenGame~\cite{intro_8} builds an agentic framework for web games. On professional engines, GameDevBench~\cite{intro_10} defines local modification tasks on Godot, AutoUE~\cite{rw_2} and UnrealLLM~\cite{rw_3} construct generation systems on Unreal Engine 5, and UniGen~\cite{rw_4} and DreamGarden~\cite{rw_5} explore multi-agent approaches on Unity and Unreal. These efforts have pushed game code generation toward professional engines, but existing work is either small in scale or relies on subjective evaluation (Table~\ref{tab:comparison}).

% \textbf{游戏与代码数据集} 从数据集规模来看，游戏领域现有的评测数据集远小于系统性研究所需的规模。GameDevBench 包含 132 个任务，OpenGame-Bench 提供 150 个 prompt，AutoUE 构建了 20 个游戏生成任务，V-GameGym 虽然包含 2,219 个样本但均为 Pygame 单文件程序。这些数据集在规模或项目复杂度上存在明显局限。在通用代码生成领域，评测粒度经历了从函数级到仓库级的持续提升：HumanEval 和 MBPP 评测函数级代码生成，ClassEval 将粒度扩展到类级，SWE-bench 和 DevEval 进一步推进到仓库级的代码修复和补全。然而，这些代码评测基准均不涉及游戏领域，且验证方式依赖单元测试，不适用于游戏这种需要运行时交互验证的场景。我们的 JamSet 从 24 万个仓库中筛选出 8,133 个经过验证的项目级游戏工程，在规模和粒度上均实现了显著跨越。
\noindent\textbf{Game and Code Datasets.}
From a dataset perspective, existing game evaluation datasets are far smaller than what systematic research requires: GameDevBench~\cite{intro_10} contains 132 tasks, OpenGame~\cite{intro_8} 150 prompts, AutoUE~\cite{rw_2} 20 tasks, and V-GameGym~\cite{intro_9,a7} 2,219 single-file samples. In general code generation, evaluation has progressed from function-level (HumanEval~\cite{intro_13,a1,a2}, MBPP~\cite{intro_14}) through class-level (ClassEval~\cite{intro_16}) to repository-level (SWE-bench~\cite{intro_17,a5,a6}, DevEval~\cite{intro_18}, RepoBench~\cite{intro_26}, CrossCodeEval~\cite{rw_10,a3,a4}, R2E-Eval~\cite{rw_11}, CodAgentBench~\cite{rw_12}). However, none address the game domain, where unit tests cannot capture runtime interactive behaviors. Our pipeline  filters 8,133 verified project-level game projects from 240K repositories.

% \textbf{评测方法论} 现有游戏评测方法可归纳为三类。第一类是手写测试脚本，以 GameDevBench 为代表，为每个任务编写定制化的测试脚本检查节点状态和游戏行为，结果精确且确定性可复现，但编写成本高，难以规模化。第二类是 VLM/LLM 自动打分，以 OpenGame 和 V-GameGym 为代表，通过视觉语言模型或语言模型对生成的游戏截图或代码进行自动评分，可以规模化但评分主观，难以精确复现。AutoUE 也采用 LLM-as-judge 对场景、玩法和视觉三个维度打分。第三类是面向传统软件的单元测试范式，以 SWE-bench 为代表，通过执行仓库自带的测试用例验证代码修复的正确性，客观且可复现，但其 input-output 验证模式无法捕捉游戏特有的运行时交互行为。我们的评测框架基于 Godot headless 引擎构建了从静态编译到运行时行为采集的确定性验证管道，兼具客观性、可复现性和可规模化。表 1 对现有工作进行了横向对比。
\noindent\textbf{Evaluation Methodology.}
Existing game evaluation methods fall into three categories. Hand-written test scripts (e.g., GameDevBench~\cite{intro_10,a15,a17}, ProxyWar~\cite{rw_8})~\cite{intro_15} are precise and reproducible but expensive to scale. VLM/LLM scoring (e.g., OpenGame~\cite{intro_8}, V-GameGym~\cite{intro_9}, AutoUE~\cite{rw_2})~\cite{intro_25} is scalable but subjective and hard to reproduce. Unit testing (e.g., SWE-bench~\cite{intro_17,a14,a12}, BigCodeBench~\cite{intro_27})~\cite{intro_18,intro_26,a10} is objective but unable to capture game-specific runtime behaviors. Our framework builds a deterministic pipeline on Godot's headless engine from compilation to runtime behavior collection, without relying on any LLM judgment or manual annotation.

\begin{figure*}[t]
\centering
\includegraphics[width=\textwidth]{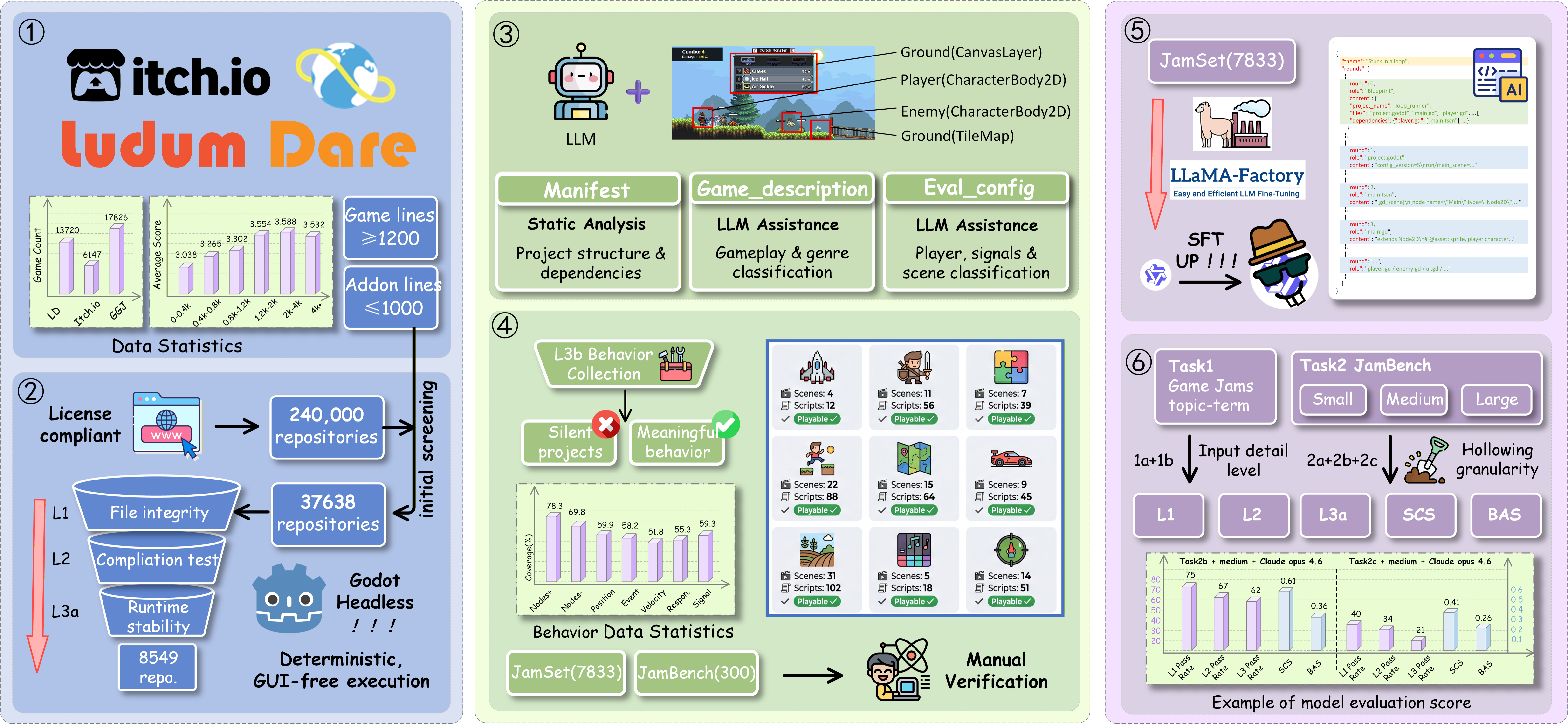}
\caption{Overview of the GameJamBench pipeline. \textcircled{1} Data collection from Game Jam platforms. \textcircled{2} Multi-level filtering via Godot headless. \textcircled{3} Structured annotation. \textcircled{4} L3b behavior collection and dataset split. \textcircled{5} Training data construction and SFT. \textcircled{6} Evaluation tasks and metrics.}
\label{fig:pipeline}
\end{figure*}

\section{Dataset Construction}

% 3.1 Game Jam 生态与数据来源
% Game Jam 是限时游戏开发竞赛，参赛者需在 48-72 小时内围绕给定主题从零构建完整游戏。这类竞赛规模庞大且参赛作品通常以开源形式共享，构成了天然的大规模真实游戏工程语料库。我们从 Ludum Dare、itch.io 和 Global Game Jam 三个平台采集数据，合并去重后共 37,588 个仓库，从中识别出 5,034 个使用 Godot 引擎的项目。Godot 的场景文件（.tscn）、脚本文件（.gd）和项目配置（.godot）均为可读文本，无二进制序列化，天然适配 LLM 的文本处理范式。为确定数据集的筛选标准，我们对 1,872 个同时具有 LD 评分和 GitHub 仓库的项目进行了相关性分析。游戏代码行数与综合评分的 Spearman ρ=0.3086（p<0.001），且 game_lines<1,200 处存在显著评分断崖。据此，我们将 game_lines≥1,200 作为质量下限，addon_lines<1,000 作为插件依赖上限，数据来源和与评分相关性见图X。
\subsection{Game Jam Ecosystem and Data Sources}
\label{sec:ecosystem}
Game Jams are time-limited game development competitions in which participants build complete games from scratch within 48--72 hours around a given theme. These events are large in scale and entries are typically shared as open-source projects, forming a natural large-scale corpus of real-world game engineering artifacts. We collect data from three major platforms: Ludum Dare, itch.io, and Global Game Jam. After merging and deduplication, we obtain 37,588 repositories, from which 5,034 Godot engine projects are identified. Godot's scene files (.tscn), script files (.gd), and project configuration (.godot) are all human-readable plain text with no binary serialization, making them naturally compatible with LLM text processing.

To establish filtering criteria for the dataset, we conduct a correlation analysis on 1,872 projects that have both Ludum Dare ratings and GitHub repositories. The Spearman correlation between game code lines (excluding third-party plugins) and overall rating is $\rho = 0.445$ ($p < 0.001$), with a significant rating cliff below 1,200 lines. Based on this analysis, we set \texttt{game\_lines} $\geq$ 1,200 as the quality threshold and \texttt{addon\_lines} $<$ 1,000 as the plugin dependency limit. Data sources and correlation results are shown in Figure~\ref{fig:godot_growth}.

\subsection{Data Collection and Filtering}
The 5,034 Game Jam projects alone are insufficient for systematic research. We conduct large-scale searches across \textbf{GitHub} and additional jam platforms including \textbf{GMTK Game Jam, Godot Wild Jam, GitHub Game Off, and Brackeys Jam}, yielding approximately 240,000 candidate repositories. We apply the criteria from Section~\ref{sec:ecosystem} as pre-filters (Godot 4.x, 2D, \texttt{game\_lines} $\geq$ 1,200, \texttt{addon\_lines} $<$ 1,000), retaining 37,638 projects. These are then passed through three levels of verification: L1 checks file integrity, L2 performs compilation verification, and L3a runs the game for 30 seconds to verify stability (technical details in Section~\ref{sec:pipeline}). From over 240,000 candidates, 8,549 projects pass all three levels. Approximately 96\% of candidates are filtered out due to missing files, compilation errors, version incompatibility, or runtime crashes, demonstrating the high noise level in open-source game repositories. Figure~\ref{fig:pipeline} illustrates the complete collection and filtering pipeline.

% 3.3 数据标注
% 为支持评测和训练数据构造，我们对上述项目进行了结构化标注：
% manifest.json 通过静态分析全自动提取，包含每个项目的脚本列表、场景树结构、输入映射和自动加载配置，以及场景跳转图。
% game_description.json 通过 LLM（Qwen3.5-397B）辅助生成，输入为 manifest.json 的信息，输出包括玩法概述、游戏类型分类。我们收集了来自 Ludum Dare、Global Game Jam、GMTK、Godot Wild Jam 和 Brackeys Jam 五个平台的 108 个真实 Game Jam 主题词，使用 sentence-transformers 对每个游戏的描述进行 embedding 匹配，最终 89 个主题词被使用。
% eval_config.json 通过 LLM 识别每个项目的玩家节点、分数与血量追踪机制、关键信号、菜单场景与游戏场景的区分等信息。这些信息用于行为采集时的精准输入策略生成（详见 4.4 节）。
% asset_annotations.json 采用 VLM 与规则混合标注。扫描 .gd、.tscn 和 .tres 中的 res:// 引用和 ext_resource 声明，图片素材通过 VLM 看图生成描述（对 AtlasTexture 按 region 裁剪后分别标注），音频和字体文件通过文件名和上下文推断。共完成 571,941个素材文件标注。注意, 所有 LLM 辅助标注仅用于数据整理和训练数据构造，四级验证管道不依赖任何 LLM 判断。

\vspace{-2mm}
\subsection{Data Annotation}

We perform structured annotation on these projects to support evaluation and training.

\noindent\textbf{manifest.json }is extracted fully automatically through static analysis, containing each project's script list, scene tree structure, input mappings, autoload configuration, and scene transition graph.

\noindent\textbf{game\_description.json} is generated with LLM assistance, taking manifest information as input and producing gameplay descriptions and genre classifications. We collect 108 real Game Jam themes from real Game Jams platforms (Ludum Dare, Global Game Jam, GMTK), and use sentence-transformers for embedding-based theme matching, retaining 89 themes.

\noindent\textbf{eval\_config.json} is generated via LLM to identify each project's player node, score and health tracking mechanisms, key signals, and the distinction between menu and gameplay scenes. This information drives precise input strategy generation during behavior collection (see Section~\ref{sec:l3b}).

\noindent\textbf{asset\_annotations.json} uses a hybrid approach combining VLM and rule-based methods. We scan \texttt{res://} references and \texttt{ext\_resource} declarations in .gd, .tscn, and .tres files. Image assets are annotated by VLM (with AtlasTexture regions cropped before annotation), while audio and font files are inferred from filenames and context. A total of 571,941 asset files are annotated. Note that all LLM-assisted annotations are used solely for data organization and training data construction; the verification pipeline involves no LLM judgment.

% 更多分析见附录。
\subsection{Dataset Statistics}
% 完成标注后，我们对全部8549个项目开展L3b行为采集（详见\ref{sec:l3b}节），通过\texttt{eval\_config}生成确定性输入策略，并采集游戏引擎运行60秒期间的运行时行为数据。其中，8133个项目可产生有效运行时行为，416个无响应项目被剔除。最终数据集包含8133个完整的2D游戏项目，依据游戏代码行数划分为三个等级：小型（1200–4000行）、中型（4000–15000行）、大型（15000行及以上），涵盖40余种游戏类型，如图所示。我们随机抽取S100/M100/L100共300个项目构成评测基准子集，每个项目均通过3–5分钟的实际游玩完成人工核验，核验通过率为100%。剩余7833个项目经逆向处理生成多轮对话训练数据，其token数量区间为小型项目2.7万至大型项目33.9万。数据集的详细统计信息与训练数据构建方式详见附录\ref{app:dataset}。

\begin{figure*}[t]
\centering
\includegraphics[width=\textwidth]{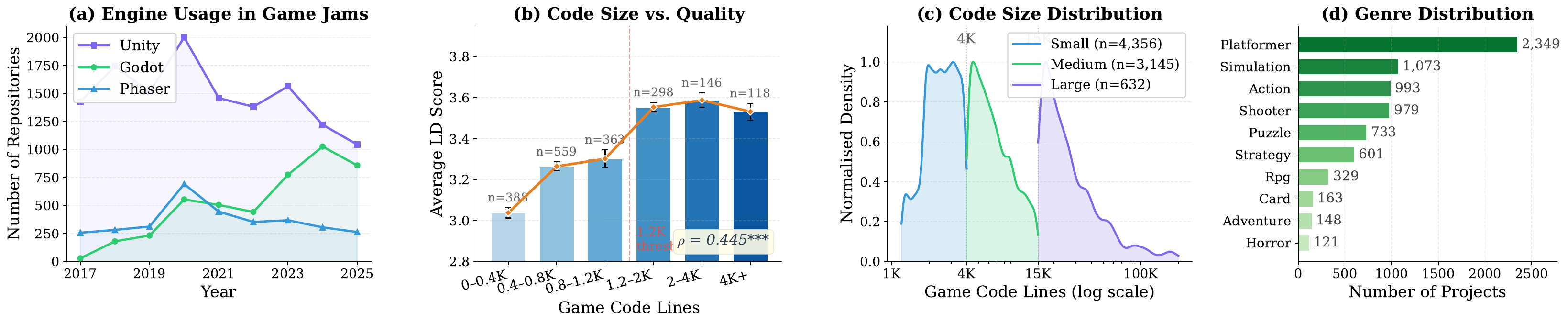}
\caption{(a) Engine usage trends in Game Jam repositories (2017--2025); (b) Code size vs. Ludum Dare rating; a quality threshold at 1.2K lines is adopted based on Spearman $\rho$=0.445 ($p$<0.001). (c) Code size distribution across three tiers with boundary cutoffs at 4K and 15K lines. (d) Top 10 genre distribution across 8,133 projects.}
\label{fig:godot_growth}
\end{figure*}

After annotation, we run L3b behavior collection on all 8,549 projects (see Section~\ref{sec:l3b}), generating deterministic input strategies from \texttt{eval\_config} and collecting runtime behavior data over 60 seconds of engine execution. Of these, 8,133 projects produce meaningful runtime behavior; 416 silent projects are excluded. These are divided into three tiers at 4K and 15K code lines: Small, Medium, and Large, spanning over 40 game genres, as shown in Figure~\ref{fig:godot_growth}c,d. We randomly sample S100/M100/L100 totaling 300 projects, each manually verified by playing for 3--5 minutes with a 100\% pass rate, both to cross-validate the reliability of our automated pipeline and to ensure benchmark quality. The remaining 7,833 projects are reverse-engineered into multi-turn dialogue training data, with token counts ranging from 21K (Small) to 197K (Large). Detailed dataset statistics and training data construction are provided in Appendix~\ref{app:dataset}.

\section{Benchmark Design}
% 4.1 任务定义
% 我们定义两类评测任务：
% Task 1：主题驱动的从零生成。 给定 Game Jam 主题词，模型需从零构建完整的 Godot 项目。其分为两个子任务：1a 仅给出主题词（如"Stuck in a loop"），需完成从创意构思到工程实现的全过程；1b 额外提供玩法描述，模型只需专注于工程实现。两者对比可分离创意规划与工程实现的能力贡献。评测形式：模型首轮输出项目蓝图，后续各轮按依赖序逐文件生成，可参考蓝图和已生成文件的摘要。测试集包含 50 个真实 Game Jam 主题词，重复三遍实验。
% Task 2：多粒度代码补全。 给定真实游戏项目，挖空部分代码后让模型补全。定义三级粒度：2a 函数级（清空 30-50% 函数体，标记为 pass # INCOMPLETE）、2b 脚本级（删除 30-50% 的 .gd 文件）、2c 全脚本级（删除所有 .gd 文件）。.tscn 场景文件完整保留。挖空使用固定种子（seed=42），所有模型使用同一份数据。评测形式：模型接收项目蓝图和保留文件的内容作为上下文，判断缺失位置之后依次输出补全内容。测试集为 300 个 benchmark 项目，每一个都有三种粒度。
\subsection{Task Definition}
\noindent\textbf{Task 1: Theme-driven generation.} Given a Game Jam theme, the model must build a complete Godot project from scratch. Task~1a provides only a theme keyword (e.g., ``Stuck in a loop''), while Task~1b additionally provides a gameplay description, isolating creative planning from engineering implementation. The model first outputs a project blueprint, then generates files in dependency order. The test set contains 50 real Game Jam themes, with experiments repeated three times.

\noindent\textbf{Task 2: Multi-granularity code completion.} Given a real game project with portions of code removed, the model must complete the missing parts. We define three granularity levels: 2a function-level (30--50\% of function bodies cleared), 2b script-level (30--50\% of .gd files removed), and 2c full-script-level (all .gd files removed). Scene files (.tscn) are fully preserved. The model receives the project blueprint and all retained file contents as context, and outputs completions sequentially. The test set consists of 300 benchmark projects, each with all three granularity levels.

\subsection{Verification and Evaluation Pipeline}
\label{sec:pipeline}

We build a four-level verification pipeline, fully deterministic and reproducible. During dataset construction, L1/L2/L3a are used for progressive filtering (Section~3.2), while L3b performs behavior collection and filters out silent projects (Section~3.4). During model evaluation, L1/L2/L3a serve as pass/fail tests, and L3b is used solely for data collection to compute BAS, not for filtering.

\noindent\textbf{L1: File integrity.} Verifies that \texttt{project.godot} exists and targets Godot 4.x, that the main scene is configured, and that all scripts and sub-scenes referenced in .tscn files are reachable. Projects with over 30\% 3D content are excluded. References within the \texttt{addons} directory are handled separately to avoid false rejections.

\noindent\textbf{L2: Compilation correctness.} Runs Godot headless compilation to catch syntax errors, type errors, and missing resource references. Checks for essential code patterns: input handling and game loop are required, while collision detection, state management, and scene transition are recommended.

\noindent\textbf{L3a: Runtime stability.} Launches the game in headless mode for 30 seconds with no input injection, verifying no crash occurs. This level filters projects that crash on startup, time out during resource loading, or are incompatible with execution.

\noindent\textbf{L3b: Runtime behavior collection.} Relies on \texttt{eval\_config} annotations to automatically generate a deterministic input strategy, then runs the game in the engine for 60 seconds while collecting multi-dimensional behavior data. Technical details are provided in Section~\ref{sec:l3b}.

\begin{table*}[t]
\centering
\small
\setlength{\tabcolsep}{8pt}
\begin{tabular}{ll ccccc ccccc}
\toprule
& & \multicolumn{5}{c}{\textbf{Task 1a (Theme Only)}} & \multicolumn{5}{c}{\textbf{Task 1b (Theme + Description)}} \\
\cmidrule(lr){3-7} \cmidrule(lr){8-12}
Cate. & Model & L1 & L2 & L3a & SCS & BAS & L1 & L2 & L3a & SCS & BAS \\
\midrule
\multirow{9}{*}{LLM}
& \cellcolor{skyblue!20}Gemini 3.1 Pro & \underline{100.0} & \underline{85.3} & \underline{78.7} & \cellcolor{red!8}0.37 & \cellcolor{red!6}0.14 & 96.0 & \underline{62.7} & 58.7 & \cellcolor{red!14}0.57 & \underline{\cellcolor{red!12}0.31} \\
& \cellcolor{skyblue!20}Claude Opus 4.6 & \underline{100.0} & 80.7 & 77.3 & \cellcolor{red!10}0.41 & \cellcolor{red!4}0.11 & 94.7 & 57.3 & 50.0 & \underline{\cellcolor{red!18}0.66} & 0.26 \\
& \cellcolor{skyblue!20}GPT-5.4 & 94.7 & 77.3 & 77.3 & \underline{\cellcolor{red!12}0.46} & \underline{\cellcolor{red!8}0.17} & 94.7 & 66.0 & \underline{60.7} & \cellcolor{red!16}0.63 & \underline{\cellcolor{red!12}0.31} \\
& \cellcolor{lightskyblue!20}DeepSeek V4 Pro & 97.3 & 82.7 & 72.7 & \cellcolor{red!4}0.28 & \cellcolor{red!2}0.09 & 93.3 & 57.3 & 52.7 & \cellcolor{red!10}0.44 & \cellcolor{red!6}0.17 \\
& \cellcolor{lightskyblue!20}Kimi K2.5 & 92.0 & 79.3 & 72.0 & \cellcolor{red!8}0.39 & \cellcolor{red!6}0.16 & 92.0 & 60.7 & 51.3 & \cellcolor{red!10}0.46 & \cellcolor{red!8}0.21 \\
& \cellcolor{lightskyblue!20}GLM-5 & 98.0 & 80.7 & 70.7 & \cellcolor{red!6}0.35 & \cellcolor{red!8}0.17 & 91.3 & 61.3 & 49.3 & \cellcolor{red!10}0.43 & \cellcolor{red!8}0.24 \\
& \cellcolor{lightskyblue!20}Qwen3.5-397B & 94.0 & 73.3 & 67.3 & \cellcolor{red!6}0.35 & \cellcolor{red!8}0.17 & 78.7 & 59.3 & 53.3 & \cellcolor{red!8}0.41 & \cellcolor{red!8}0.22 \\
& \cellcolor{lightskyblue!20}Qwen3.5-27B & 78.7 & 66.0 & 58.7 & \cellcolor{red!4}0.27 & \cellcolor{red!2}0.09 & 68.7 & 51.3 & 29.3 & \cellcolor{red!6}0.33 & \cellcolor{red!4}0.11 \\
& \cellcolor{sftgold!20}Qwen3.5-27B-SFT & 88.7 & 70.7 & 62.0 & \cellcolor{red!6}0.34 & \cellcolor{red!8}0.21 & 86.0 & 61.3 & 34.7 & \cellcolor{red!8}0.41 & \cellcolor{red!8}0.23 \\
\midrule
\multirow{5}{*}{Agent}
& \cellcolor{skyblue!20}Claude Opus 4.6 & \underline{100.0} & \underline{92.7} & \underline{82.7} & \underline{\cellcolor{red!10}0.42} & \cellcolor{red!4}0.13 & \underline{100.0} & \underline{82.7} & \underline{80.0} & \underline{\cellcolor{red!16}0.62} & \underline{\cellcolor{red!10}0.28} \\
& \cellcolor{lightskyblue!20}DeepSeek V4 Pro & \underline{100.0} & 93.3 & 84.0 & \cellcolor{red!6}0.34 & \cellcolor{red!4}0.12 & \underline{100.0} & 80.0 & 76.7 & \cellcolor{red!12}0.49 & \cellcolor{red!6}0.14 \\
& \cellcolor{lightskyblue!20}Kimi K2.5 & \underline{100.0} & 90.0 & 81.3 & \cellcolor{red!6}0.31 & \underline{\cellcolor{red!8}0.19} & \underline{100.0} & 78.0 & 77.3 & \cellcolor{red!8}0.37 & \cellcolor{red!6}0.17 \\
& \cellcolor{lightskyblue!20}GLM-5 & \underline{100.0} & 91.3 & 78.7 & \cellcolor{red!8}0.37 & \cellcolor{red!8}0.17 & \underline{100.0} & 77.3 & 73.3 & \cellcolor{red!8}0.41 & \cellcolor{red!4}0.11 \\
& \cellcolor{lightskyblue!20}Qwen3.5-397B & \underline{100.0} & 90.0 & 80.7 & \cellcolor{red!6}0.31 & \cellcolor{red!6}0.14 & \underline{100.0} & 70.7 & 68.7 & \cellcolor{red!10}0.43 & \cellcolor{red!6}0.16 \\
\bottomrule
\end{tabular}
\caption{Task~1 from-scratch generation results. L1/L2/L3a are pass rates (\%). SCS and BAS are mean over 3 runs.}
\label{tab:task1}
\end{table*}

% 4.3 评测指标
% 我们分别从静态结构和运行时行为两个维度刻画模型的游戏工程能力：
% 结构完整度评分（SCS）。 通过静态分析衡量代码框架的结构完整程度，包含 8 个维度：脚本数、场景数、总代码行数、函数密度、节点密度、非空函数比例、信号使用数和 input_action 数量。每维度计算生成值与参照值的比值（截断至 1.0），取平均：

% $$SCS = \frac{1}{N}\sum_{i=1}^{N} \min\left(\frac{v_i}{r_i}, 1.0\right)$$

% 其中 $v_i$ 为生成项目第 $i$ 维度的值，$r_i$ 为参照值，$N=8$。Task 1 中 $r_i$ 取数据集中同类型游戏的平均值，Task 2 中 $r_i$ 取原版项目的对应值。当 $r_i = 0$ 且 $v_i = 0$ 时得分为 1.0。

% 行为对齐评分（BAS）。 通过 L3b 采集结果衡量运行时行为的正确程度，包含 7 个维度：位置变化（position_changes）、速度活跃帧（velocity_active）、分数变化（score_changes）、血量变化（health_changes）、信号触发数（signal_trigger_count）、节点增删数（node_changes）和场景切换数（scene_changes）。各维度相似度定义为：

% $$s_j^{num} = 1 - \frac{|b_j - b_j^{ref}|}{\max(b_j, b_j^{ref})}, \quad s^{sig} = \frac{|S_{gen} \cap S_{ref}|}{|S_{gen} \cup S_{ref}|}$$

% 其中 $b_j$ 和 $b_j^{ref}$ 分别为生成项目和参照项目第 $j$ 维度的行为值，$S_{gen}$ 和 $S_{ref}$ 为各自触发的信号类型集合。BAS 总分为：

% $$BAS = \frac{1}{|\mathcal{D}_{num}| + 1}\left(\sum_{j \in \mathcal{D}_{num}} s_j^{num} + s^{sig}\right)$$

% Task 2 中 $b_j^{ref}$ 为原版项目的行为数据，采用上述对称相似度；Task 1 中参照值为数据集均值，数值维度改用 $s_j = \min(b_j / b_j^{ref}, 1.0)$。当双方均为零时得分为 1.0。

% 模型输出通过 L1、L2、L3a、SCS 和 BAS 五个指标进行评测。其中 L1/L2/L3a 为通过性验证，SCS 和 BAS 为量化评分。通过编译但 SCS 低，说明模型以极简代码投机通过编译；SCS 高但 BAS 低，说明静态结构完整但动态交互很差。二者互补，共同刻画模型的游戏工程能力。

\subsection{Evaluation Metrics}

\begin{table}[t]
\centering
\small
\setlength{\tabcolsep}{4pt}
\resizebox{\columnwidth}{!}{%
\begin{tabular}{lc lc}
\toprule
\textbf{SCS Dimension} & \textbf{Ref.} & \textbf{BAS Dimension} & \textbf{Ref.} \\
\midrule
script\_count       & 26.42  & nodes\_added          & 4720.31   \\
scene\_count        & 22.22  & nodes\_removed        & 4411.50   \\
input\_action       & 6.53   & position\_changes     & 131.18   \\
function\_count     & 128.62 & event\_count          & 33.38   \\
node\_count         & 201.55 & velocity\_active      & 102.17   \\
non\_empty\_ratio   & 0.98   & responsive\_actions   & 4.73   \\
signal\_usage       & 58.02  & signal\_triggers      & 4.0   \\
\bottomrule
\end{tabular}%
}
\caption{Dataset reference values for Task~1 evaluation. SCS references are dataset means. BAS references are means from projects with rich runtime behavior. BAS dimensions are selected with $>$50\% non-zero coverage.}
\label{tab:reference}
\end{table}

We evaluate game engineering along two dimensions: static structure and runtime behavior.

\noindent \textbf{Structural Completeness Score (SCS).} SCS measures the structural completeness of the generated code framework through static analysis across 7 dimensions: script count, scene count, input action count, function count, node count, non-empty function ratio, and signal usage. These dimensions capture the core structural aspects of Godot game projects. SCS is defined as:
\begin{equation}
\resizebox{0.53\columnwidth}{!}{$\displaystyle
    SCS = \frac{1}{N}\sum_{i=1}^{N} \min\left(\frac{v_i}{v_i^{ref}},\ 1.0\right)
$}
\end{equation}
where $v_i$ is the generated value on dimension $i$, $v_i^{ref}$ is the reference value, and $N=7$. For Task~1, $v_i^{ref}$ is the same-genre dataset mean; for Task~2, $v_i^{ref}$ is the original project's value. When $v_i = v_i^{ref} = 0$, the score is 1.0.

\noindent \textbf{Behavioral Alignment Score (BAS).} We select behavioral dimensions with non-zero coverage exceeding 50\% across the dataset, yielding 7 numeric dimensions (nodes added, nodes removed, position changes, event count, velocity active frames, responsive actions) and 1 set dimension (signal trigger type overlap). Detailed coverage statistics by genre are provided in Appendix~\ref{app:bas_coverage}. The per-dimension similarity is:
\begin{equation}
\resizebox{0.78\columnwidth}{!}{$\displaystyle
    s_j^{num}\!=\!1\!-\!\frac{|b_j\!-\!b_j^{ref}|}{\max(b_j,b_j^{ref})},\quad s^{sig}\!=\!\frac{|S_{gen}\!\cap\!S_{ref}|}{|S_{gen}\!\cup\!S_{ref}|}
$}
\end{equation}
When both $b_j = b_j^{ref} = 0$, $s_j^{num} = 1.0$; when both $S_{gen} = S_{ref} = \emptyset$, $s^{sig} = 1.0$.
The overall BAS is:
% \begin{equation}
% \small BAS = \frac{1}{|\mathcal{D}_{num}|+1}\left(\sum_{j \in \mathcal{D}_{num}} s_j^{num} + s^{sig}\right)
% \end{equation}
\begin{equation}
\resizebox{0.7\columnwidth}{!}{$\displaystyle
    BAS = \frac{1}{|\mathcal{D}_{num}|+1}\left(\sum_{j \in \mathcal{D}_{num}} s_j^{num} + s^{sig}\right)
$}
\end{equation}
For Task~2, references are the original project's behavior data. For Task~1, references are dataset means (Table~\ref{tab:reference}) with capped ratio $s_j = \min(b_j / b_j^{ref},\ 1.0)$. Only non-zero dimensions contribute; if fewer than 2 are non-zero, one zero-valued dimension is included.

Model outputs are evaluated using five metrics: L1, L2, and L3a as pass/fail verification, and SCS and BAS as quantitative scores. The two scores are complementary: low SCS with passing compilation indicates minimal code, while high SCS with low BAS indicates structurally complete but behaviorally incorrect output.

\subsection{L3b Behavior Collection}
\label{sec:l3b}

The core challenge is to perform meaningful interactions and collect behavior data for thousands of diverse games in a headless environment while ensuring full determinism. We address this with a three-layer architecture: (1) offline LLM preprocessing generates \texttt{eval\_config} from project structure, identifying player nodes, key signals, and scene classification; (2) rule-based input strategy generation produces deterministic action sequences from \texttt{eval\_config} with no LLM calls; (3) runtime execution injects inputs and collects 7-dimension behavior data over 60 seconds, with menu navigation in the first 10 seconds and gameplay interaction in the remaining 50. Notably, the goal of behavior collection is not to play the game to completion, but to trigger as many active nodes as possible, producing rich behavioral signals for comparison between generated and reference projects. The entire pipeline guarantees determinism: the same \texttt{eval\_config} always produces the same input strategy and runtime behavior. Technical details and the full algorithm are provided in Appendix~\ref{app:technical}.

\begin{table*}[t]
\centering
\setlength{\tabcolsep}{6pt}
\scriptsize
\resizebox{\textwidth}{!}{%
\begin{tabular}{ll ccccc ccccc ccccc}
\toprule
& & \multicolumn{5}{c}{\textbf{Level 2a (Function)}} & \multicolumn{5}{c}{\textbf{Level 2b (Script)}} & \multicolumn{5}{c}{\textbf{Level 2c (Full-script)}} \\
\cmidrule(lr){3-7} \cmidrule(lr){8-12} \cmidrule(lr){13-17}
Tier & Model & L1 & L2 & L3a & SCS & BAS & L1 & L2 & L3a & SCS & BAS & L1 & L2 & L3a & SCS & BAS \\
\midrule
\multirow{9}{*}{{Small}}
& \cellcolor{skyblue!20}Gemini 3.1 Pro & 97 & 88 & \underline{88} & \cellcolor{red!12}0.94 & \cellcolor{red!18}0.69 & \underline{100} & \underline{96} & \underline{93} & \cellcolor{red!20}0.83 & \underline{\cellcolor{red!20}0.64} & \underline{96} & \underline{85} & \underline{77} & \cellcolor{red!12}0.62 & \underline{\cellcolor{red!20}0.51} \\
& \cellcolor{skyblue!20}Claude Opus 4.6 & \underline{100} & \underline{90} & 84 & \cellcolor{red!6}0.92 & \underline{\cellcolor{red!20}0.71} & \underline{100} & 92 & 88 & \cellcolor{red!16}0.79 & \cellcolor{red!10}0.52 & 93 & 81 & \underline{77} & \cellcolor{red!16}0.65 & \cellcolor{red!4}0.37 \\
& \cellcolor{skyblue!20}GPT-5.4 & 96 & 84 & 81 & \underline{\cellcolor{red!20}0.97} & \cellcolor{red!14}0.64 & \underline{100} & 91 & 88 & \underline{\cellcolor{red!20}0.85} & \cellcolor{red!4}0.46 & 90 & 78 & 75 & \underline{\cellcolor{red!20}0.70} & \cellcolor{red!2}0.35 \\
& \cellcolor{lightskyblue!20}DeepSeek V4 Pro & 92 & 85 & 81 & \cellcolor{red!10}0.93 & \cellcolor{red!6}0.57 & 99 & 93 & \underline{93} & \cellcolor{red!12}0.73 & \cellcolor{red!10}0.51 & 92 & 80 & 72 & \cellcolor{red!8}0.58 & \cellcolor{red!10}0.42 \\
& \cellcolor{lightskyblue!20}Kimi K2.5 & 93 & 82 & 80 & \cellcolor{red!10}0.93 & \cellcolor{red!10}0.60 & \underline{100} & 95 & 90 & \cellcolor{red!14}0.76 & \cellcolor{red!16}0.59 & 90 & 79 & 74 & \cellcolor{red!8}0.57 & \cellcolor{red!12}0.44 \\
& \cellcolor{lightskyblue!20}GLM-5 & 90 & 83 & 79 & 0.90 & \cellcolor{red!6}0.57 & 97 & 89 & 83 & \cellcolor{red!12}0.74 & \cellcolor{red!12}0.54 & 89 & 81 & 70 & \cellcolor{red!12}0.62 & \cellcolor{red!10}0.41 \\
& \cellcolor{lightskyblue!20}Qwen3.5-397B & 94 & 86 & 81 & \cellcolor{red!14}0.95 & \cellcolor{red!12}0.62 & 92 & 90 & 84 & \cellcolor{red!14}0.77 & \cellcolor{red!12}0.54 & 87 & 75 & 71 & \cellcolor{red!16}0.66 & \cellcolor{red!6}0.38 \\
& \cellcolor{lightskyblue!20}Qwen3.5-27B & 83 & 73 & 70 & \cellcolor{red!6}0.92 & 0.51 & 88 & 76 & 72 & 0.56 & 0.41 & 78 & 69 & 62 & 0.50 & 0.33 \\
& \cellcolor{sftgold!20}Qwen3.5-27B-SFT & 86 & 80 & 80 & \cellcolor{red!10}0.93 & \cellcolor{red!10}0.60 & 90 & 85 & 81 & \cellcolor{red!12}0.73 & \cellcolor{red!14}0.57 & 82 & 73 & 67 & \cellcolor{red!12}0.61 & \cellcolor{red!4}0.37 \\
\midrule
\multirow{9}{*}{{Medium}}
& \cellcolor{skyblue!20}Gemini 3.1 Pro & \underline{61} & 50 & 32 & \cellcolor{red!20}0.80 & \cellcolor{red!4}0.46 & 72 & 59 & 49 & \cellcolor{red!16}0.54 & \cellcolor{red!20}0.35 & 43 & 32 & 19 & \cellcolor{red!16}0.37 & \cellcolor{red!18}0.23 \\
& \cellcolor{skyblue!20}Claude Opus 4.6 & 56 & 53 & \underline{44} & \underline{\cellcolor{red!20}0.82} & \underline{\cellcolor{red!20}0.51} & \underline{75} & \underline{67} & \underline{62} & \underline{\cellcolor{red!20}0.61} & \underline{\cellcolor{red!20}0.36} & 40 & \underline{34} & \underline{23} & \cellcolor{red!20}0.41 & \underline{\cellcolor{red!20}0.26} \\
& \cellcolor{skyblue!20}GPT-5.4 & 59 & 48 & 38 & \cellcolor{red!16}0.76 & \cellcolor{red!8}0.47 & 68 & 61 & 56 & \cellcolor{red!18}0.56 & \underline{\cellcolor{red!20}0.36} & \underline{47} & 33 & 20 & \cellcolor{red!18}0.39 & \cellcolor{red!12}0.19 \\
& \cellcolor{lightskyblue!20}DeepSeek V4 Pro & 55 & 44 & 34 & \cellcolor{red!12}0.68 & 0.45 & 69 & 64 & 58 & \cellcolor{red!16}0.53 & \cellcolor{red!14}0.31 & 39 & 32 & 17 & \underline{\cellcolor{red!20}0.43} & \cellcolor{red!14}0.20 \\
& \cellcolor{lightskyblue!20}Kimi K2.5 & \underline{61} & \underline{54} & 43 & \cellcolor{red!12}0.69 & \underline{\cellcolor{red!20}0.51} & 65 & 57 & 53 & \cellcolor{red!14}0.49 & \cellcolor{red!16}0.33 & 42 & 30 & 16 & \cellcolor{red!16}0.36 & \cellcolor{red!18}0.24 \\
& \cellcolor{lightskyblue!20}GLM-5 & 47 & 42 & 37 & \cellcolor{red!10}0.66 & \cellcolor{red!8}0.47 & 67 & 60 & 45 & \cellcolor{red!12}0.47 & \cellcolor{red!10}0.27 & 40 & 26 & \underline{23} & \cellcolor{red!12}0.32 & \cellcolor{red!12}0.19 \\
& \cellcolor{lightskyblue!20}Qwen3.5-397B & 56 & 48 & 41 & \cellcolor{red!10}0.67 & \cellcolor{red!18}0.50 & 65 & 53 & 53 & \cellcolor{red!14}0.51 & \cellcolor{red!12}0.29 & 37 & 31 & 19 & \cellcolor{red!10}0.29 & \cellcolor{red!20}0.25 \\
& \cellcolor{lightskyblue!20}Qwen3.5-27B & 32 & 29 & 17 & 0.52 & \cellcolor{red!4}0.46 & 48 & 46 & 23 & 0.28 & 0.19 & 25 & 13 & 6 & 0.15 & 0.08 \\
& \cellcolor{sftgold!20}Qwen3.5-27B-SFT & 43 & 31 & 26 & \cellcolor{red!6}0.61 & 0.45 & 52 & 44 & 38 & \cellcolor{red!8}0.40 & \cellcolor{red!8}0.26 & 33 & 20 & 14 & \cellcolor{red!6}0.22 & \cellcolor{red!10}0.17 \\
\midrule
\multirow{9}{*}{{Large}}
& \cellcolor{skyblue!20}Gemini 3.1 Pro & \underline{32} & 14 & 7 & \cellcolor{red!14}0.53 & \cellcolor{red!10}0.50 & 39 & 18 & 13 & \cellcolor{red!18}0.43 & \cellcolor{red!20}0.30 & 9 & 4 & 1 & \underline{\cellcolor{red!20}0.29} & 0.01 \\
& \cellcolor{skyblue!20}Claude Opus 4.6 & 28 & \underline{17} & \underline{12} & \cellcolor{red!10}0.50 & \cellcolor{red!14}0.52 & \underline{41} & \underline{22} & \underline{16} & \underline{\cellcolor{red!20}0.45} & \underline{\cellcolor{red!20}0.31} & \underline{15} & \underline{9} & \underline{2} & \cellcolor{red!14}0.23 & \underline{\cellcolor{red!20}0.06} \\
& \cellcolor{skyblue!20}GPT-5.4 & 30 & 14 & 7 & \cellcolor{red!6}0.48 & \underline{\cellcolor{red!20}0.56} & 35 & 14 & 11 & \cellcolor{red!18}0.43 & \cellcolor{red!16}0.26 & 11 & 4 & \underline{2} & 0.08 & \cellcolor{red!4}0.02 \\
& \cellcolor{lightskyblue!20}DeepSeek V4 Pro & 24 & 10 & 4 & \underline{\cellcolor{red!20}0.58} & \cellcolor{red!8}0.49 & 27 & 9 & 9 & \cellcolor{red!16}0.41 & \cellcolor{red!20}0.29 & 10 & 6 & 0 & 0.00 & 0.00 \\
& \cellcolor{lightskyblue!20}Kimi K2.5 & 26 & 11 & 5 & \cellcolor{red!4}0.46 & \cellcolor{red!8}0.48 & 32 & 17 & 12 & \cellcolor{red!12}0.38 & \cellcolor{red!16}0.24 & 8 & 5 & 0 & 0.00 & 0.00 \\
& \cellcolor{lightskyblue!20}GLM-5 & 23 & 13 & 9 & 0.44 & 0.44 & 29 & \underline{22} & 14 & \cellcolor{red!14}0.39 & \cellcolor{red!16}0.26 & 8 & 4 & 1 & \cellcolor{red!12}0.21 & \cellcolor{red!12}0.04 \\
& \cellcolor{lightskyblue!20}Qwen3.5-397B & 19 & 11 & 5 & \cellcolor{red!18}0.56 & \cellcolor{red!8}0.48 & 25 & 17 & 13 & \cellcolor{red!14}0.40 & \cellcolor{red!10}0.18 & 6 & 3 & 0 & 0.00 & 0.00 \\
& \cellcolor{lightskyblue!20}Qwen3.5-27B & 3 & 0 & 0 & 0.00 & 0.00 & 11 & 4 & 2 & 0.27 & 0.04 & 0 & 0 & 0 & 0.00 & 0.00 \\
& \cellcolor{sftgold!20}Qwen3.5-27B-SFT & 12 & 4 & 2 & \cellcolor{red!2}0.45 & \cellcolor{red!12}0.51 & 19 & 13 & 8 & \cellcolor{red!8}0.33 & \cellcolor{red!6}0.11 & 3 & 0 & 0 & 0.00 & 0.00 \\
\bottomrule
\end{tabular}
}
\caption{Task~2 Direct LLM results by Tier and granularity. L1/L2/L3a: pass rates (\%). SCS/BAS: scores in $[0,1]$.}
\label{tab:task2_llm}
\end{table*}

\section{Experiments}
% 5.1 实验设置
% 我们设置两类评测实验。第一类为 Direct LLM 实验，直接使用语言模型进行生成，包括模型（Claude Opus 4、GPT-5.4、Gemini 3.1 Pro、Kimi K2.5、DeepSeek V4 Pro、Qwen3.5-397B、GLM-5、Qwen3.5-27B）以及一个微调模型（Qwen3.5-27B-SFT），其基于 Qwen3.5-27B使用Jam Set进行lora微调，训练超参数详见附录 B。第二类为 Code Agent 实验，以 Claude Code 作为 agent 框架，分别驱动 Claude、DeepSeek、Kimi、GLM 和 Qwen 五个模型执行迭代调试。所有模型在 Task 1 和 Task 2（见 4.1 节）的完整测试集上进行评测，并报告 L1、L2、L3a、SCS 和 BAS 五个指标（见 4.3 节）。
\subsection{Experimental Setup}
We conduct two categories of evaluation experiments. The first is \textbf{Direct LLM}, where LLMs generate outputs directly. This category includes 9 models (Claude Opus 4.6, GPT-5.4, Gemini 3.1 Pro, Kimi K2.5, DeepSeek V4 Pro, Qwen3.5-397B, GLM-5, Qwen3.5-27B) and one fine-tuned model (Qwen3.5-27B-SFT), which is LoRA‑fine‑tuned on Qwen3.5‑27B using Jam Set. Training hyperparameters are detailed in Appendix~B. The second is \textbf{Code Agent}, using Claude Code as the agent framework to drive LLMs through iterative debugging. All models are evaluated on the full test sets of Task~1 and Task~2 (Section~4.1), reporting L1, L2, L3a, SCS, and BAS (Section~4.3). Token length statistics for training data and evaluation prompts are provided in Appendix~\ref{app:dataset}.

% 5.2 LLM 直接生成实验
% 表 X 报告了九个语言模型在 Task 1 上的表现。从零构建游戏对所有模型在编译通过率方面表现良好：编译通过率为--。但值得注意的是，模型的SCS和BAS表现远远低于编译通过率，表明其生成了能够编译但结构极简的项目，说明模型倾向于“偷懒”只保证游戏基础框架的搭建。给定玩法描述后（1b），多数模型的SCS和BAS出现提升，但是编译通过率同步下滑，证明模型的设计复杂度和工程能力之间存在权衡。Qwen3.5-27B-SFT 相比原版在编译通过率和 SCS 上均有提升，验证了训练数据的有效性。

% 表 X 报告了 Task 2 的结果，按 Tier（Small/Medium/Large）和粒度（2a/2b/2c）分解。从规模变化的角度分析：项目规模是模型工程能力的硬瓶颈，所有模型在 Task 2 中从 Small 到 Large 均呈现断崖式下降，这一点观察和软件工程领域研究(swe-bench等)的发现一致。更详细的趋势分析见第 6 章。
\subsection{Direct LLM Results}
Table~\ref{tab:task1} reports the performance of 9 LLMs on Task~1. Building games from scratch yields reasonable pass rates: L3a averages 70.7\%. However, SCS and BAS scores are far lower, indicating that models tend to produce compilable but structurally minimal projects. When provided with gameplay descriptions (Task~1b), most models show improved SCS and BAS, but compilation pass rates decline, revealing a trade-off between design complexity and engineering capability. Qwen3.5-27B-SFT shows improvements over its base version in both compilation rate and SCS, validating the effectiveness of the training data.

Table~\ref{tab:task2_llm} reports Task~2 results, broken down by Tiers and granularity (2a/2b/2c). Project scale is a  bottleneck: all models exhibit a sharp performance cliff from Small to Large, consistent with findings in software engineering research. More detailed trend analysis is provided in Section~6.
\subsection{Code Agent Results}
Table~\ref{tab:task1} reports Code Agent performance on Task~1, and Table~\ref{tab:task2_full} reports results on Task~2. The Agent mode improves pass rates across the board: on Task~1a, Claude+Agent raises L3a from 77.3\% to 82.7\%, and DeepSeek+Agent from 72.7\% to 84.0\%. However, SCS and BAS show minimal change: Claude+Agent achieves SCS of 0.42 (LLM: 0.41) and BAS of 0.13 (LLM: 0.11). More detailed analysis is provided in Section~6.

\begin{figure*}[t]
\centering
\includegraphics[width=\textwidth]{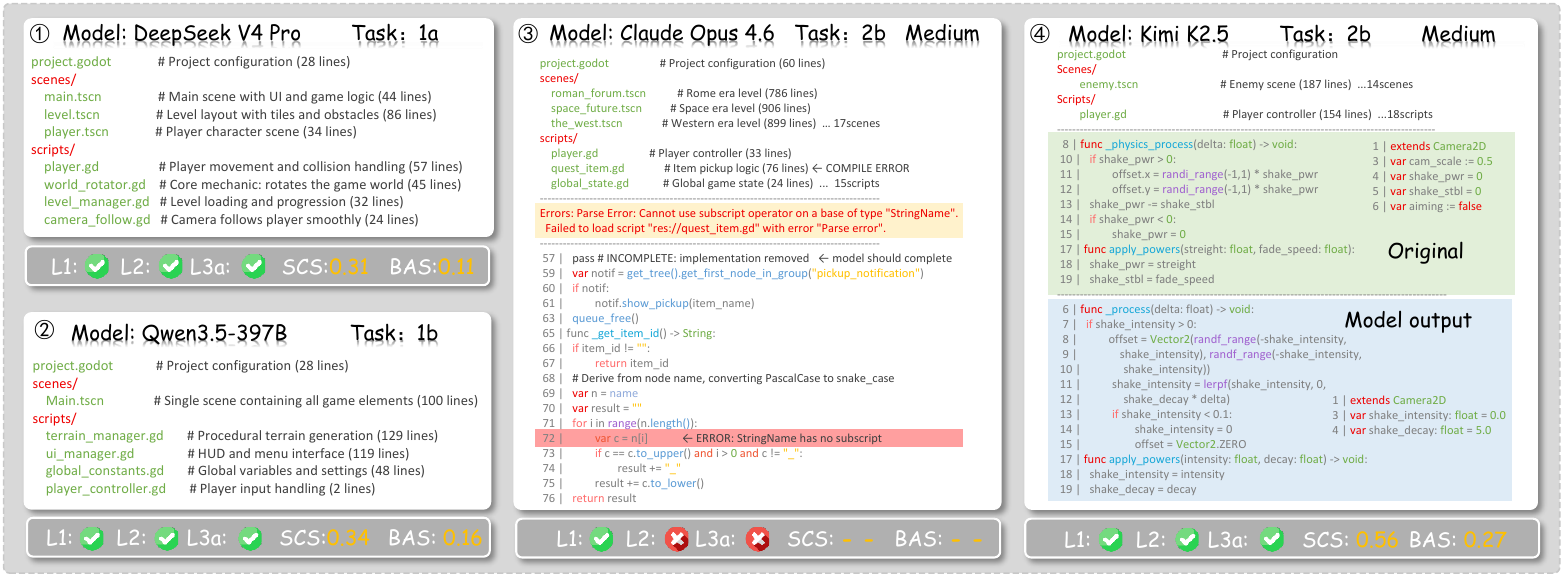}
\caption{Representative case studies. Case~1 (\ding{192}\ding{193}): Passing verification but with minimal functionality. Case~2(\ding{194}): Compilation failure. Case~3 (\ding{195}): semantically drifted implementation breaking cross-file dependencies.}
\label{fig:case_study}
\end{figure*}

\section{Analysis}
% 6.1 核心洞察
% 基于第五章的实验结果，我们总结以下核心发现。
\subsection{Core Insights}

\noindent\textbf{Function-level completion demands more contextual precision.}
A counter-intuitive pattern emerges: despite removing more code, script-level completion (2b) achieves higher compilation pass rates than function-level (2a). Function-level completion requires precise restoration consistent with retained code in interfaces and logic, where a single incorrect function can break compilation. Script-level completion allows rewriting with internally consistent logic, free from existing code constraints. Interestingly, SCS and BAS show the opposite trend: 2a scores higher than 2b, as more original code is preserved. This suggests that completion granularity does not follow a simple difficulty progression, and function-level completion demands higher contextual precision. ClassEval~\cite{intro_16} reports similar findings in a different domain.

% Code Agent 的帮助局限于编译层面。 引入 Agent 后 L2 通过率平均提升超过 40 个百分点，但 SCS 和 BAS 均未随之改善。在 Task 1a 中，Agent 将编译率大幅拉高，但项目的结构完整度和运行时行为质量与纯 LLM 基本持平；在 Task 1b 中，模型本身因获得玩法描述已能构建相对完整的项目结构，Agent 同样仅提升了编译率而未改善 SCS 和 BAS。Task 2 中实验也呈现一致的趋势。这说明 Agent 的迭代调试能力集中在修复编译错误和文件引用等语法问题，对游戏项目的整体质量无明显贡献。与此形成对比的是，Task 1 中 1b 相比 1a 的 SCS 显著提升表明，给模型提供更好的设计输入比给模型提供语法修复工具更为有效，创意设计到工程实现的转化能力是当前的制约因素之一。

\noindent \textbf{Code Agent improvements are limited to compilation.}
Agents raise L3a pass rates by over 30 percentage points on average, yet SCS and BAS show no corresponding improvement. In Task~1, Agents improve compilation rates, but structural completeness and runtime behavior remain on par with direct LLM outputs, indicating that compilability alone is not a sufficient measure of game quality; Task~2 exhibits a consistent pattern (Appendix~\ref{app:agent_results}). In contrast, the SCS gain from Task~1a to 1b shows that detailed gameplay planning can effectively guide the model to generate more complete game projects.

% 工程范式差异。 我们对比模型生成的项目与真实项目的工程特征，发现两者存在系统性差异。最显著的表现是输入映射：数据集中真实项目平均定义 6.53 个 input_action，倾向于通过抽象接口处理玩家输入；而模型生成的项目中 input_action_count 大部分为 0，完全依赖硬编码按键事件。类似的差异还体现在全局状态管理上：76.3% 的真实项目使用 autoload 全局脚本管理游戏状态，而模型倾向于将逻辑集中在单个脚本中。这些差异表明，当前模型学到的是"能运行的代码片段"而非"规范的游戏工程实践"，它们缺乏对专业游戏引擎项目级组织范式的理解。值得注意的是，经过 JamSet 训练数据微调后的 Qwen3.5-27B-SFT 在上述指标上出现了显著变化：其生成项目的平均 input_action_count 从 0.08 提升至 3.44，autoload 使用率从 55.1% 提升至 77.1%，场景跳转逻辑的使用率也从 18.0% 提升至 49.4%。这表明 JamSet 的训练数据不仅提升了模型的编译通过率和游戏内容质量，更在工程组织范式上引导模型向人类开发者的实践方式靠拢。更多分析见附录 C。

\noindent \textbf{Engineering paradigm gap.}
Model-generated projects exhibit systematic differences from real projects. Most notably, real projects define an average of 6.9 input actions via abstract interfaces, while model-generated projects rely entirely on hard-coded key events (\texttt{input\_action\_count} near zero). Similarly, 76.3\% of real projects use autoload scripts for global state management, while models concentrate logic in single scripts. These differences suggest that models learn to produce code that runs rather than proper game engineering practices. After fine-tuning on JamSet, Qwen3.5-27B-SFT shows significant improvement: \texttt{input\_action\_count} rises from 0.08 to 3.44, autoload usage from 55.1\% to 77.1\%, and scene transition usage from 18\% to 49.4\%, demonstrating that domain-specific training data guides models toward human-like engineering practices.

\subsection{Case Study}
% 我们通过三个典型案例进一步阐释上述发现（图 X）。
% 案例一：投机型输出。 DeepSeek V4 Pro 和 Qwen3.5-397B 在 Task 1a 中均通过了全部三级验证，但SCS得分较低。值得注意的是 Qwen 的项目包含了 terrain_manager（129 行）和 ui_manager（119 行）等看似复杂的模块，但核心的 player_controller.gd 仅有 2 行代码，模型并未完善游戏核心逻辑。这印证了第五章的观察：模型倾向于生成能通过编译的最小可行结构，而非功能完整的游戏项目。同时也验证了 SCS 和 BAS 作为编译验证的补充指标的必要性：仅凭编译通过率无法区分"能跑的空壳"和"真正的游戏"。
% 案例二：专有知识的不足。 Claude Opus 4.6 在 Task 2b 中展现了正确的算法设计能力：PascalCase 到 snake_case 的转换逻辑完全正确, 但是在代码语法正确性上出现问题，导致程序崩溃。这说明：模型虽然有强大的编程范式，但缺乏对特定引擎 API 的精确掌握。同样解释了模型与人类的工程范式差异：人类开发者通过引擎文档和实践积累了引擎特定知识，而模型的训练数据中此类知识严重不足，印证了我们数据集重要性。
% 案例三：跨文件脆弱性。 Kimi K2.5 在 Task 2b 中为 camera.gd 生成了功能等价的相机抖动实现，代码本身可正常运行。然而其重写了全部变量名和函数签名，破坏了其他脚本通过 camera.cam_scale 和 camera.apply_powers() 进行的跨文件调用。这说明：模型能理解单个模块的功能意图，却在面对较大规模项目时难以维护项目级的接口统一，这是函数级补全（2a）和脚本级补全（2b）呈现不同失败模式的根本原因。
We present three typical cases(Figure~\ref{fig:case_study}).

\noindent\textbf{Case 1: Shortcut outputs.} DeepSeek V4 Pro and Qwen3.5-397B both pass three verification levels on Task~1a, yet achieve low SCS scores. Notably, Qwen's project contains seemingly complex modules such as terrain\_manager (129 lines) and ui\_manager (119 lines), but the core player\_controller.gd has only 2 lines of code, with the model failing to implement any actual game logic. This confirms that compilation pass rates alone cannot distinguish between a runnable shell and a real game, validating SCS and BAS as necessary complementary metrics.

\noindent\textbf{Case 2: Lack of engine-specific knowledge.} Claude Opus 4.6 demonstrates correct algorithm design in Task~2b, implementing a fully valid PascalCase-to-snake\_case conversion, yet the code fails to compile due to a syntax issue with engine-specific types. The algorithm itself is correct in general-purpose programming, but GDScript imposes type constraints that differ from standard languages. Engine-specific Godot code constitutes a tiny fraction of existing training data, making it difficult for models to master these conventions. JamSet filters out substantial noise and organizes projects into a training-friendly structured format. Fine-tuning experiments show that training on JamSet notably improves model output quality, confirming the value of curated, structured domain data for bridging this knowledge gap.

\noindent\textbf{Case 3: Cross-file semantic drift.} Kimi K2.5 generates a functionally equivalent camera shake implementation in Task~2b that compiles and runs correctly. However, the rewritten code omits key state variables that other scripts rely on for camera zoom and aiming, causing silent behavioral degradation. This explains the gap between SCS (0.56) and BAS (0.27): structure appears complete but runtime interactions are incomplete, showing that models can replicate individual module intent but struggle to preserve cross-file implicit interfaces.

\section{Conclusion}
% 本文构建了首个专业游戏引擎上的大规模项目级游戏代码框架数据集 JamSet 和评测基准 JamBench。我们利用 Godot 引擎的 headless 模式设计了确定性四级验证管道，从超过 24 万个开源仓库中筛选出 8,133 个经过验证的完整 2D 游戏项目，并定义了主题驱动的从零生成和多粒度代码补全两类评测任务，配合结构完整度评分（SCS）和行为对齐评分（BAS）构成多层评测框架。对前沿模型的系统评测揭示了三个核心发现：项目规模是模型工程能力的硬瓶颈；Code Agent 的提升局限于编译层面，创意设计到工程实现的转化能力仍然是制约因素之一；模型与人类开发者在游戏工程范式上存在系统性差异，而基于 JamSet 的微调可以有效缩小这一差距，验证了数据集的有效性。

We present JamSet and JamBench, the first project-level game code framework dataset and benchmark on a professional game engine. Using a deterministic four-level verification pipeline on Godot's headless mode, we distill 8,133 verified 2D game projects from over 240,000 repositories, and define from-scratch generation and multi-granularity completion tasks evaluated by SCS and BAS. Evaluation of frontier models reveals three findings: project scale is a hard capability bottleneck; Code Agent improvements are limited to compilation; and models exhibit systematic engineering paradigm gaps from human developers, which fine-tuning on JamSet effectively narrows.

\section*{Scope and Limitations}
This work focuses on evaluating code frameworks rather than complete games with art and audio assets. Assessing such assets is inherently subjective and often dominates perceived game quality, undermining the reproducibility we require. Moreover, while asset generation and gameplay design have received considerable attention in prior work, code framework generation on professional engines remains  unexplored. Since a well-structured code framework can support arbitrary asset integration and flexible design extensions, we argue that its verification is a meaningful and self-contained objective.

This work has the following limitations. First, the dataset focuses on the Godot game engine, which is growing rapidly as an open-source engine, but Unity, Unreal, and other engines still hold significant market share. The current dataset does not cover these engines, and we plan to continuously expand the application boundary in future work. Second, this work focuses on dataset construction and benchmark design. The exploration of model training serves only as preliminary validation of data effectiveness, without thorough ablation experiments. In the future, we plan to augment the dataset through project variants, repair projects that did not pass verification to expand the data scale, and train a domain-specific expert model for game engineering on this basis.

\section*{Ethics Statement}
\paragraph{Data Sources and Licensing.}
All game projects used in this work are sourced from publicly available open-source repositories across platforms including Ludum Dare, itch.io, Global Game Jam, and GitHub. Only projects with open-source licenses are included. The dataset contains no personal privacy information, and all project metadata have been stripped of developer identity markers.

\paragraph{Model Usage.}
Language models and vision-language models used during data annotation serve solely to generate structured annotations (gameplay descriptions, evaluation configurations, asset descriptions) and are not involved in any verification step. The four-level verification pipeline is fully automated and relies on no model judgment.

\paragraph{Open-Source Commitment.}
We will publicly release the complete dataset, evaluation framework, and tool code to facilitate research in game engineering automation.



\bibliographystyle{abbrv}
\bibliography{custom}
\clearpage
\appendix
\raggedbottom

\section*{Appendix}

\section*{Appendix Overview}
\startcontents[sections]
\printcontents[sections]{l}{1}{\setcounter{tocdepth}{2}}

\section{Additional Related Work}
\label{app:code_comparison}
In the main text we primarily compare with game-domain benchmarks (Table~\ref{tab:comparison}). To further clarify our positioning in the broader code engineering evaluation landscape, we provide a detailed comparison with general-purpose code benchmarks here.

\textbf{SWE-bench}~\cite{intro_17} collects 2,294 test tasks from 12 Python repositories (with a training set of 19,000 tasks across 37 repositories), requiring models to generate patches that fix real GitHub issues, verified through existing test suites. In terms of dataset construction, SWE-bench relies on GitHub issues and corresponding pull requests, with each task requiring developer-written test cases and manual environment configuration; our data is automatically filtered from 240,000 repositories without per-project manual setup. In terms of evaluation, SWE-bench executes unit tests and reports pass/fail rates; we collect runtime behavior through the game engine without relying on pre-written test cases. In terms of task type, SWE-bench focuses on local repair; we extend beyond local repair to include from-scratch generation and multi-granularity completion.

\textbf{DevEval}~\cite{intro_18} is manually annotated by 13 developers, containing 1,874 code generation samples from 117 Python repositories that simulate developers writing code in real repositories, verified through unit tests. In terms of construction, DevEval relies entirely on manual annotation, which is costly and difficult to scale; our pipeline is fully automated, producing a dataset over 4$\times$ larger. In terms of evaluation dimensions, DevEval reports Pass@1 and other pass/fail metrics; we additionally provide SCS and BAS as continuous indicators that capture cases where code compiles but lacks functionality. Furthermore, DevEval focuses on function-level and file-level generation, while we evaluate complete project generation and completion.

\textbf{BigCodeBench}~\cite{intro_27} contains 1,140 function-level code generation tasks emphasizing diverse third-party library calls and complex instructions, verified through execution of test cases. In terms of construction, BigCodeBench tasks are manually designed with hand-written test cases; our test data comes from real game projects without requiring hand-written tests. In terms of granularity, BigCodeBench operates at the function level (single function bodies); we operate at the project level. BigCodeBench similarly reports pass/fail for function correctness.

\textbf{ProjectEval}~\cite{a2} contains 20 tasks with 284 test cases, constructed by LLM generation with human review. Evaluation verifies functional correctness by simulating user interactions through test cases. In terms of construction, ProjectEval tasks are LLM-generated then human-reviewed, with scale limited by review cost (20 tasks); our data comes from real Game Jam projects, automatically filtered to 8,133 projects through a deterministic pipeline without per-project manual review. In terms of evaluation, ProjectEval relies on pre-written test scripts that simulate user interaction, requiring manual test design for each task; our evaluation requires no per-project test cases. Table~\ref{tab:code_comparison} provides a detailed comparison.

\textbf{DevBench}~\cite{a27} is a code completion benchmark containing 1,800 instances across 6 programming languages and 6 task categories. Its evaluation combines three methods: functional correctness checks, similarity metrics, and LLM-judge scoring. In terms of construction, DevBench extracts task types from real developer telemetry, then generates instances synthetically with human review, making the process semi-automated; our data comes directly from real game projects. In terms of evaluation, DevBench incorporates LLM-judge as part of its assessment, providing richer quality dimensions but at the cost of reproducibility, as LLM scoring may vary across models or over time. In terms of granularity, DevBench focuses on code completion level (line/function); we evaluate complete project generation and completion.

Among mainstream benchmarks in both game and general code domains, our work offers differentiated advantages in scale, automation, and evaluation dimensions.

\begin{table*}[h]
\centering
\small
\resizebox{\textwidth}{!}{%
\begin{tabular}{lcccccccc}
\toprule
\textbf{Benchmark} & \textbf{Domain} & \textbf{Scale} & \textbf{Granularity} & \textbf{Construction} & \textbf{Verification} & \textbf{Det.} & \textbf{Scal.} & \textbf{Runtime} \\
\midrule
SWE-bench & Python & 2,294 & Repo repair & Issue+PR & Unit test & \cmark & \xmark & \xmark \\
DevEval & Python & 1,874 & Func/File & Manual & Unit test & \cmark & \xmark & \xmark \\
DevBench & Multi-lang & 1,800 & Line/Func & Telemetry+Synth & LLM judge & \xmark & \cmark & \xmark \\
BigCodeBench & Python & 1,140 & Function & Manual & Execution & \cmark & \xmark & \xmark \\
ProjectEval & Python & 20 & Project & LLM+Human & Interaction sim. & \cmark & \xmark & \cmark \\
\midrule
\rowcolor{blue!4} Ours & Godot & 8,133 & Project & Auto pipeline & Engine headless & \cmark & \cmark & \cmark \\
\bottomrule
\end{tabular}
}
\caption{Comparison with code engineering benchmarks. Det. = Deterministic. Scal. = Scalable (no per-task manual annotation or test writing). Runtime = Evaluates runtime behavior.}
\label{tab:code_comparison}
\end{table*}

\section{Additional Experiment Results}
\label{app:agent_results}

\begin{table*}[h!]
\centering
\setlength{\tabcolsep}{3.5pt}
\small
\resizebox{\textwidth}{!}{%
\begin{tabular}{lll ccccc ccccc ccccc}
\toprule
& & & \multicolumn{5}{c}{Level 2a (Function)} & \multicolumn{5}{c}{Level 2b (Script)} & \multicolumn{5}{c}{Level 2c (Full-script)} \\
\cmidrule(lr){4-8} \cmidrule(lr){9-13} \cmidrule(lr){14-18}
Tier & Type & Model & L1 & L2 & L3a & SCS & BAS & L1 & L2 & L3a & SCS & BAS & L1 & L2 & L3a & SCS & BAS \\
\midrule
\multirow{12}{*}{{Small}}
& \multirow{9}{*}{{LLM}}
& \cellcolor{skyblue!20}Gemini 3.1 Pro   & 97 & 88 & 88 & 0.94 & 0.69 & 100 & 96 & 93 & 0.83 & 0.64 & 96 & 85 & 77 & 0.62 & 0.51 \\
& & \cellcolor{skyblue!20}Claude Opus 4.6    & 100 & 90 & 84 & 0.92 & 0.71 & 100 & 92 & 88 & 0.79 & 0.52 & 93 & 81 & 77 & 0.65 & 0.37 \\
& & \cellcolor{skyblue!20}GPT-5.4          & 96 & 84 & 81 & 0.97 & 0.64 & 100 & 91 & 88 & 0.85 & 0.46 & 90 & 78 & 75 & 0.70 & 0.35 \\
& & \cellcolor{lightskyblue!20}DeepSeek V4 Pro  & 92 & 85 & 81 & 0.93 & 0.57 & 99 & 93 & 93 & 0.73 & 0.51 & 92 & 80 & 72 & 0.58 & 0.42 \\
& & \cellcolor{lightskyblue!20}Kimi K2.5        & 93 & 82 & 80 & 0.93 & 0.60 & 100 & 95 & 90 & 0.76 & 0.59 & 90 & 79 & 74 & 0.57 & 0.44 \\
& & \cellcolor{lightskyblue!20}GLM-5       & 90 & 83 & 79 & 0.90 & 0.57 & 97 & 89 & 83 & 0.74 & 0.54 & 89 & 81 & 70 & 0.62 & 0.41 \\
& & \cellcolor{lightskyblue!20}Qwen3.5-397B& 94 & 86 & 81 & 0.95 & 0.62 & 92 & 90 & 84 & 0.77 & 0.54 & 87 & 75 & 71 & 0.66 & 0.38 \\
& & \cellcolor{lightskyblue!20}Qwen3.5-27B & 83 & 73 & 70 & 0.92 & 0.51 & 88 & 76 & 72 & 0.56 & 0.41 & 78 & 69 & 62 & 0.50 & 0.33 \\
& & \cellcolor{sftgold!20}Qwen3.5-27B-SFT  & 86 & 80 & 80 & 0.93 & 0.60 & 90 & 85 & 81 & 0.73 & 0.57 & 82 & 73 & 67 & 0.61 & 0.37 \\
\cmidrule(lr){2-18}
& \multirow{3}{*}{{Agent}}
& \cellcolor{skyblue!20}Claude Opus 4.6   & 100 & 96 & 90 & 0.93 & 0.72 & 100 & 98 & 94 & 0.80 & 0.53 & 100 & 92 & 85 & 0.66 & 0.38 \\
& & \cellcolor{lightskyblue!20}DeepSeek V4 Pro  & 100 & 94 & 88 & 0.94 & 0.58 & 100 & 97 & 95 & 0.74 & 0.52 & 100 & 90 & 80 & 0.59 & 0.43 \\
& & \cellcolor{lightskyblue!20}Qwen3.5-397B      & 96 & 91 & 85 & 0.93 & 0.66 & 100 & 96 & 92 & 0.78 & 0.56 & 99 & 87 & 77 & 0.52 & 0.34 \\
\midrule
\multirow{12}{*}{{Medium}}
& \multirow{9}{*}{{LLM}}
& \cellcolor{skyblue!20}Gemini 3.1 Pro   & 61 & 50 & 32 & 0.80 & 0.46 & 72 & 59 & 49 & 0.54 & 0.35 & 43 & 32 & 19 & 0.37 & 0.23 \\
& & \cellcolor{skyblue!20}Claude Opus 4.6    & 56 & 53 & 44 & 0.82 & 0.51 & 75 & 67 & 62 & 0.61 & 0.36 & 40 & 34 & 23 & 0.41 & 0.26 \\
& & \cellcolor{skyblue!20}GPT-5.4          & 59 & 48 & 38 & 0.76 & 0.47 & 68 & 61 & 56 & 0.56 & 0.36 & 47 & 33 & 20 & 0.39 & 0.19 \\
& & \cellcolor{lightskyblue!20}DeepSeek V4 Pro  & 55 & 44 & 34 & 0.68 & 0.45 & 69 & 64 & 58 & 0.53 & 0.31 & 39 & 32 & 17 & 0.43 & 0.20 \\
& & \cellcolor{lightskyblue!20}Kimi K2.5        & 61 & 54 & 43 & 0.69 & 0.51 & 65 & 57 & 53 & 0.49 & 0.33 & 42 & 30 & 16 & 0.36 & 0.24 \\
& & \cellcolor{lightskyblue!20}GLM-5       & 47 & 42 & 37 & 0.66 & 0.47 & 67 & 60 & 45 & 0.47 & 0.27 & 40 & 26 & 23 & 0.32 & 0.19 \\
& & \cellcolor{lightskyblue!20}Qwen3.5-397B& 56 & 48 & 41 & 0.67 & 0.50 & 65 & 53 & 53 & 0.51 & 0.29 & 37 & 31 & 19 & 0.29 & 0.25 \\
& & \cellcolor{lightskyblue!20}Qwen3.5-27B & 32 & 29 & 17 & 0.52 & 0.46 & 48 & 46 & 23 & 0.28 & 0.19 & 25 & 13 & 6 & 0.15 & 0.08 \\
& & \cellcolor{sftgold!20}Qwen3.5-27B-SFT  & 43 & 31 & 26 & 0.61 & 0.45 & 52 & 44 & 38 & 0.40 & 0.26 & 33 & 20 & 14 & 0.22 & 0.17 \\
\cmidrule(lr){2-18}
& \multirow{3}{*}{{Agent}}
& \cellcolor{skyblue!20}Claude Opus 4.6   & 88 & 76 & 70 & 0.80 & 0.52 & 88 & 80 & 76 & 0.62 & 0.37 & 62 & 57 & 52 & 0.42 & 0.27 \\
& & \cellcolor{lightskyblue!20}DeepSeek V4 Pro  & 85 & 77 & 68 & 0.69 & 0.46 & 89 & 78 & 73 & 0.54 & 0.32 & 58 & 58 & 56 & 0.44 & 0.21 \\
& & \cellcolor{lightskyblue!20}Qwen3.5-397B      & 78 & 74 & 70 & 0.68 & 0.47 & 84 & 78 & 74 & 0.50 & 0.27 & 62 & 58 & 54 & 0.37 & 0.29 \\
\midrule
\multirow{12}{*}{{Large}}
& \multirow{9}{*}{{LLM}}
& \cellcolor{skyblue!20}Gemini 3.1 Pro   & 32 & 14 & 7 & 0.53 & 0.50 & 39 & 18 & 13 & 0.43 & 0.30 & 9 & 4 & 1 & 0.29 & 0.01 \\
& & \cellcolor{skyblue!20}Claude Opus 4.6    & 28 & 17 & 12 & 0.50 & 0.52 & 41 & 22 & 16 & 0.45 & 0.31 & 15 & 9 & 2 & 0.23 & 0.06 \\
& & \cellcolor{skyblue!20}GPT-5.4          & 30 & 14 & 7 & 0.48 & 0.56 & 35 & 14 & 11 & 0.43 & 0.26 & 11 & 4 & 2 & 0.08 & 0.02 \\
& & \cellcolor{lightskyblue!20}DeepSeek V4 Pro  & 24 & 10 & 4 & 0.58 & 0.49 & 27 & 9 & 9 & 0.41 & 0.29 & 10 & 6 & 0 & 0.00 & 0.00 \\
& & \cellcolor{lightskyblue!20}Kimi K2.5        & 26 & 11 & 5 & 0.46 & 0.48 & 32 & 17 & 12 & 0.38 & 0.24 & 8 & 5 & 0 & 0.00 & 0.00 \\
& & \cellcolor{lightskyblue!20}GLM-5       & 23 & 13 & 9 & 0.44 & 0.44 & 29 & 22 & 14 & 0.39 & 0.26 & 8 & 4 & 1 & 0.21 & 0.04 \\
& & \cellcolor{lightskyblue!20}Qwen3.5-397B& 19 & 11 & 5 & 0.56 & 0.48 & 25 & 17 & 13 & 0.40 & 0.18 & 6 & 3 & 0 & 0.00 & 0.00 \\
& & \cellcolor{lightskyblue!20}Qwen3.5-27B & 3 & 0 & 0 & 0.00 & 0.00 & 11 & 4 & 2 & 0.27 & 0.04 & 0 & 0 & 0 & 0.00 & 0.00 \\
& & \cellcolor{sftgold!20}Qwen3.5-27B-SFT  & 12 & 4 & 2 & 0.45 & 0.51 & 19 & 13 & 8 & 0.33 & 0.11 & 3 & 0 & 0 & 0.00 & 0.00 \\
\cmidrule(lr){2-18}
& \multirow{3}{*}{{Agent}}
& \cellcolor{lightskyblue!20}Claude Opus 4.6   & 46 & 37 & 28 & 0.51 & 0.53 & 50 & 40 & 35 & 0.46 & 0.32 & 33 & 23 & 21 & 0.14 & 0.13 \\
& & \cellcolor{green!10}DeepSeek V4 Pro  & 42 & 37 & 31 & 0.59 & 0.44 & 50 & 38 & 33 & 0.41 & 0.30 & 27 & 19 & 13 & 0.11 & 0.09 \\
& & \cellcolor{lightskyblue!20}Qwen3.5-397B      & 40 & 31 & 26 & 0.48 & 0.51 & 46 & 37 & 30 & 0.44 & 0.25 & 28 & 20 & 17 & 0.13 & 0.07 \\
\bottomrule
\end{tabular}
}
\caption{Task~2 results for Direct LLM and Code Agent. L1/L2/L3a: pass rates (\%). SCS/BAS: scores in $[0,1]$.}
\label{tab:task2_full}
\end{table*}

Table~\ref{tab:task2_full} presents the complete Task~2 results for both Direct LLM and Code Agent experiments. Code Agent experiments use Claude Code as the agent framework, with ccswitch replacing the underlying model to drive Claude, DeepSeek, and Qwen3.5-397B through iterative debugging.

From the perspective of completion granularity, Task~2 exhibits the same counter-intuitive pattern: script-level completion (2b) achieves higher compilation pass rates than function-level (2a), regardless of whether LLM or Code Agent is used. In the Small tier, Claude LLM achieves 88\% L3a on 2b versus 84\% on 2a; Claude+Agent achieves 94\% on 2b versus 90\% on 2a. This trend is more pronounced in the Medium and Large tiers, further validating the finding in Section~6: function-level completion demands higher contextual precision than script-level completion, consistent with the granularity effect reported by ClassEval~\cite{intro_16} in class-level code generation.

From the perspective of Agent improvement, Agents significantly boost compilation pass rates across all tiers and granularities. In the Medium tier 2b, Claude's L3a rises from 62\% to 76\%, DeepSeek from 58\% to 73\%, and Qwen from 53\% to 74\%. However, SCS and BAS remain largely unchanged: Claude+Agent achieves 0.80 SCS on Medium 2a (LLM: 0.82) and 0.52 BAS (LLM: 0.51). This is fully consistent with Task~1 findings: Agent debugging focuses on fixing compilation errors and file references at the syntactic level, with no meaningful improvement in structural completeness or runtime behavioral quality. AgentCoder~\cite{a12} reports similar findings in general code generation: multi-agent iterative testing and optimization effectively improve test pass rates, but improvements are concentrated at the surface correctness level in scenarios requiring deep semantic understanding. Our results extend this observation to the game engineering domain: compilability is merely the minimum quality threshold, while scene structure design, cross-file logic coordination, and runtime interactive behavior are the true quality indicators. Current Agent repair mechanisms focus on eliminating compilation errors without reaching these higher-level engineering quality dimensions.

We therefore suggest that future Code Agent development should incorporate domain-specific knowledge into the iterative workflow, such as game engine engineering conventions, scene organization paradigms, and cross-file interface contracts, rather than checking only compilability at each iteration. When Agent feedback signals expand from ``does it compile'' to ``does it follow domain engineering practices,'' improvements in project quality can move beyond syntactic repair.

\section{Dataset and Training Details}
\label{app:dataset}

% --- Table C: Extends Distribution ---
\begin{table*}[h]
\centering
\small
\begin{tabular}{lrl}
\toprule
\textbf{Base Class} & \textbf{Count} & \textbf{Typical Role} \\
\midrule
Node2D           & 22,777 & 2D game objects \\
Node             & 21,786 & General logic / managers \\
Control          & 14,536 & UI elements \\
Area2D           & 10,660 & Collision / trigger zones \\
CharacterBody2D  & 8,155  & Player / NPC characters \\
Resource         & 4,647  & Data resources \\
CanvasLayer      & 4,622  & UI layers / HUD \\
Node3D           & 4,091  & 3D objects (mixed projects) \\
RefCounted       & 2,861  & Pure logic classes \\
Button           & 2,489  & Button widgets \\
\bottomrule
\end{tabular}
\caption{Top 10 script base classes (\texttt{extends}) across all projects, reflecting the structural composition of 2D game development in Godot.}
\label{tab:extends_dist}
\end{table*}

\subsection{Training Data Construction}

Each of the 7,833 training projects is reverse-engineered into a multi-turn dialogue sample that simulates building a game from scratch: the first turn generates a project blueprint from a theme, and subsequent turns generate files in dependency-topological order. The following processing rules are applied:

\noindent\textbf{GDScript files (.gd):} Fully preserved with injected \texttt{@asset} semantic descriptions for referenced art and audio assets.

\noindent\textbf{Scene files (.tscn):} Complete node trees and properties retained; engine-generated \texttt{uid}s and non-semantic binary data removed.

\noindent\textbf{Resource files (.tres):} Retained if under 10K tokens.

\noindent\textbf{Excluded:} \texttt{.import} and \texttt{.cfg} files.
The resulting training data ranges from a median of 21K tokens for Small projects to 197K for Large. A complete example is provided in Appendix~\ref{app:example}. See Figure~\ref{fig:testset_overview}e for the token length statistics of the training data.

\subsection{Training Hyperparameters}
\label{app:training}
Table~\ref{tab:hyperparams} lists the hyperparameters used for fine-tuning Qwen3.5-27B-SFT. We use LoRA with rank 8 on all linear layers, trained for 3 epochs on JamSet-Small and JamSet-Medium (7,300 samples). Large-tier projects exceed the 32K token cutoff and are excluded from training.

\begin{table*}[h]
\centering
\small
\begin{tabular}{ll}
\toprule
\textbf{Hyperparameter} & \textbf{Value} \\
\midrule
Base model & Qwen3.5-27B \\
Fine-tuning method & LoRA \\
LoRA rank & 8 \\
LoRA target & All linear layers \\
DeepSpeed stage & ZeRO-3 \\
Attention & FlashAttention-2 \\
Hardware & 8 $\times$ NVIDIA H200 (141GB) \\
Precision & FP8 + BF16 \\
Learning rate & 1e-4 \\
LR scheduler & Cosine \\
Warmup ratio & 0.1 \\
Epochs & 3 \\
Cutoff length & 32,768  \\
Per-device batch size & 1 \\
Gradient accumulation & 2 \\
Effective batch size & 16 \\
Gradient checkpointing & Enabled \\
Training data & JamSet-Small + JamSet-Medium \\
 & (4,258 + 3,042 = 7,300 samples) \\
Training data format & ShareGPT multi-turn dialogue \\
\bottomrule
\end{tabular}
\caption{Fine-tuning hyperparameters for Qwen3.5-27B-SFT.}
\label{tab:hyperparams}
\end{table*}

\subsection{BAS Dimension Coverage by Genre}
\label{app:bas_coverage}
Table~\ref{tab:bas_coverage} reports the non-zero coverage rate of each BAS dimension across game genres. Platformer games exhibit the highest average non-zero dimensions (4.8), driven by rich physics interactions, while card games show the lowest (2.5) due to minimal spatial movement. The overall average of 3.9 non-zero dimensions per project confirms that every game exhibits meaningful behavior across multiple dimensions.

\begin{table*}[h]
\centering
\small
\begin{tabular}{lrrrrrrrrc}
\toprule
\textbf{Genre} & \textbf{N} & \textbf{nodes+} & \textbf{nodes--} & \textbf{pos} & \textbf{event} & \textbf{vel} & \textbf{resp} & \textbf{signal} & \textbf{avg} \\
\midrule
platformer  & 2,349 & 74 & 70 & 85 & 62 & 92 & 79 & 60 & 4.8 \\
simulation  & 1,073 & 80 & 63 & 37 & 49 & 35 & 33 & 72 & 3.0 \\
action      & 993   & 75 & 69 & 70 & 61 & 94 & 63 & 76 & 4.3 \\
shooter     & 979   & 79 & 72 & 62 & 60 & 88 & 67 & 83 & 4.2 \\
puzzle      & 733   & 83 & 76 & 44 & 60 & 14 & 32 & 52 & 3.3 \\
strategy    & 601   & 84 & 69 & 31 & 52 &  7 & 30 & 24 & 3.0 \\
rpg         & 329   & 80 & 65 & 58 & 55 & 34 & 51 & 72 & 3.8 \\
card        & 163   & 83 & 67 & 23 & 50 & 11 & 11 & 62 & 2.5 \\
adventure   & 148   & 78 & 76 & 39 & 57 & 82 & 34 & 65 & 3.2 \\
horror      & 121   & 83 & 75 & 44 & 61 & 61 & 45 & 27 & 3.7 \\
...         &   &  & &  &  &  &  & &  \\
\midrule
All         & 8,133 & 78 & 70 & 60 & 58 & 51 & 55 & 59 & 3.9 \\
\bottomrule
\end{tabular}
\caption{Non-zero coverage rate (\%) of each BAS dimension across game genres. ``avg'' is the average number of non-zero dimensions per project. All numeric dimensions exceed 50\% overall coverage except velocity (39\%) and signal triggers (34\%); the latter is included as a set-based dimension capturing Godot's signal-driven architecture.}
\label{tab:bas_coverage}
\end{table*}

\subsection{Benchmark and Evaluation Statistics}
\begin{figure*}[t]
\centering
\includegraphics[width=\textwidth]{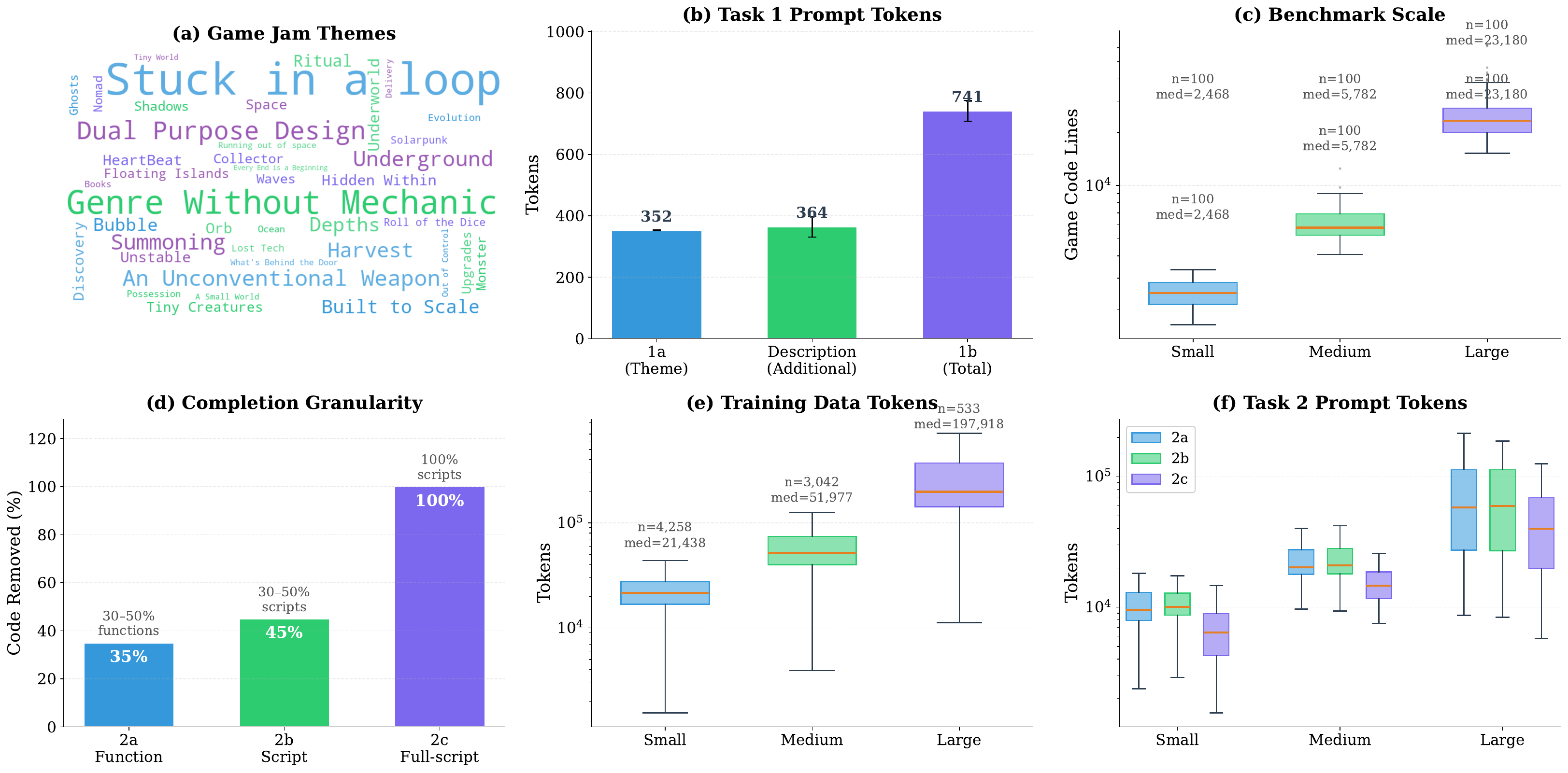}
\caption{Benchmark and evaluation statistics. (a) Game Jam theme word cloud for Task~1. (b) First-round prompt token distribution. (c) Benchmark project scale by tier. (d) Code removal ratio by completion granularity.}
\label{fig:testset_overview}
\end{figure*}

Figure~\ref{fig:testset_overview} summarizes the benchmark test set. (a) Word cloud of 50 Game Jam themes used in Task~1, sourced from 5 platforms with 89 themes matched to dataset projects. (b) First-round prompt token counts: Task~1a averages 352 tokens (theme only), while Task~1b averages 741 tokens with the additional gameplay description contributing 364 tokens. (c) Benchmark project scale distribution across three tiers, with median code lines of 2,468 (Small), 5,782 (Medium), and 23,180 (Large). (d) Code removal ratios for Task~2 completion granularities: function-level (2a) removes 30--50\% of functions, script-level (2b) removes 30--50\% of scripts, and full-script (2c) removes 100\% of script content.

\section{L3b Technical Details}
\label{app:technical}

\subsection{Three-Layer Architecture}

\noindent\textbf{Layer 1: Offline preprocessing.} Structural information is extracted from \texttt{manifest.json} (script extends and function signatures, scene node trees, input mappings) and processed by an LLM to generate \texttt{eval\_config.json}. This identifies player node type and path, score and health tracking mechanisms, key game signals (e.g., \texttt{enemy\_died}, \texttt{coin\_collected}), menu versus gameplay scene classification, and expected behavior patterns (e.g., jumping and movement for platformers, shooting for shooters). This is a one-time offline annotation with fixed results. The LLM prompt template is provided in Appendix~\ref{app:prompts}.

\noindent\textbf{Layer 2: Input strategy generation.} From \texttt{eval\_config}'s expected behaviors, input actions, and game type, \texttt{input\_strategy.json} is generated through pure rule-based transformation with no LLM calls. It contains three components: input mode (keyboard/mouse/mixed), a deterministic action sequence (mapping expected behaviors to concrete actions), and a menu navigation strategy (targeting gameplay scenes identified in \texttt{eval\_config}).

\noindent\textbf{Layer 3: Runtime deterministic execution.} As shown in Algorithm~\ref{alg:l3b}, execution proceeds in two phases. The menu navigation phase (first 10 seconds) scans Button nodes, reads their positions, and attempts to bypass menus through a three-level fallback: emit \texttt{pressed} signal, mouse click at \texttt{Button.global\_position}, keyboard Enter/Space. Upon detecting a scene change to the target gameplay scene, execution advances to the next phase. The gameplay input phase (remaining 50 seconds) executes the deterministic action sequence according to the input mode, collecting all seven BAS dimensions throughout.

\subsection{Determinism Guarantees}

The same \texttt{eval\_config} always produces the same \texttt{input\_strategy}, and runtime execution follows a fixed order. Three safeguards prevent pause-related disruptions: the collection script's \texttt{process\_mode} is set to \texttt{ALWAYS} (unaffected by game pausing), \texttt{SceneTree.paused} is checked and cleared every frame, and dangerous actions (pause, escape, quit) are excluded from input sequences.

\begin{algorithm2e}[h]
    \caption{L3b Deterministic Behavior Collection}
    \label{alg:l3b}
    \footnotesize
    \SetKwInOut{Input}{Input}
    \SetKwInOut{Output}{Output}
    \Input{eval\_config $E$, input\_strategy $S = \{\text{mode, actions, menu\_targets}\}$}
    \Output{event\_stream $B = \{\text{pos, vel, sig, score, health, node, scene}\}$}
    \tcp{Phase 1: Menu Navigation ($t \in [0, T_{menu}]$)}
    $\text{buttons} \leftarrow \{x \in \text{SceneTree} \mid x.\text{type} = \text{Button}\}$\;
    Sort buttons by keyword priority (play $>$ start $>$ begin $>$ ...)\;
    \For{each button $x$ in buttons}{
        \For{$f$ in \{emit\_pressed, mouse\_click($x$.pos), key\_inject(Enter)\}}{
            $f(x)$\;
            \If{current\_scene $\in E$.gameplay\_scenes}{
                \textbf{goto} Phase 2\;
            }
        }
    }
    \tcp{Phase 2: Gameplay Interaction ($t \in [T_{menu}, T_{max}]$)}
    \For{$t = 0, 1, \ldots, T_{max}$}{
        $a \leftarrow S.\text{actions}[t \bmod |S.\text{actions}|]$\;
        \uIf{$S.\text{mode} = \text{keyboard}$}{
            inject(ActionEvent($a$), KeyEvent(keymap($a$)))\;
        }
        \uElseIf{$S.\text{mode} = \text{mouse}$}{
            $\text{targets} \leftarrow \{x \in \text{SceneTree} \mid \text{is\_interactive}(x)\}$\;
            inject(MouseClick(targets[$t \bmod |\text{targets}|$].pos))\;
        }
        \Else{
            inject(ActionEvent($a$))\;
            \If{$t \bmod N = 0$}{
                inject(MouseClick(next\_target.pos))\;
            }
        }
        $B[t] \leftarrow \text{snapshot}(\text{pos, vel, sig, score, health, nodes, scenes})$\;
        \If{SceneTree.paused}{
            SceneTree.paused $\leftarrow$ \textbf{false}\tcp*{Unpause immediately}
        }
    }
    \KwRet{$B$}
\end{algorithm2e}

\subsection{Input Strategy Rule Mapping}

The input strategy generator maps game types identified in \texttt{eval\_config} to concrete action sequences through a deterministic rule table. Table~\ref{tab:input_mapping} shows the mapping for common game types. For each game type, the generator selects an input mode and constructs a cyclic action sequence that covers the game's expected interactions. The action sequence length is fixed at 600 frames (10 seconds at 60 FPS), repeating cyclically for the full 50-second gameplay phase. When \texttt{eval\_config} specifies custom input actions (e.g., \texttt{move\_left}, \texttt{jump}), these override the default mappings.

\begin{table*}[t]
\centering
\small
\begin{tabular}{llp{5cm}}
\toprule
\textbf{Game Type} & \textbf{Mode} & \textbf{Action Sequence} \\
\midrule
platformer & keyboard & left/right movement (60\%), jump (25\%), attack (15\%), direction alternation period: 2--3s \\
shooter & mixed & WASD movement (50\%), mouse clicks at interactive targets (30\%), attack action (20\%) \\
puzzle & mouse & click interactive nodes in scene tree order, drag gestures between clickable targets \\
strategy & mouse & click UI elements and interactive nodes, periodic keyboard confirm (Enter/Space) \\
rpg & keyboard & directional movement (40\%), interact (30\%), attack (20\%), menu navigation (10\%) \\
card & mouse & click card-like nodes, drag between positions, periodic confirm clicks \\
other & mixed & WASD movement + periodic mouse clicks at scene center and interactive nodes \\
\bottomrule
\end{tabular}
\caption{Default input strategy mapping by game type. Custom input actions from \texttt{eval\_config} override these defaults when available.}
\label{tab:input_mapping}
\end{table*}

\subsection{Menu Navigation Challenges}

Menu bypass is a significant technical challenge because Game Jam games have no standardized menu structure. Some games start directly in gameplay, others have a single ``Play'' button, and many feature multi-layered menus with settings, credits, and level selection screens. We identified three categories of menu structures in the dataset:

\noindent\textbf{No menu} (31.2\%): The main scene is the gameplay scene itself. L3b detects this when \texttt{eval\_config.menu\_scenes} is empty or the main scene appears in \texttt{gameplay\_scenes}, and skips directly to the input phase.
 
\noindent\textbf{Simple menu} (52.4\%): A single screen with Button nodes labeled ``Play'', ``Start'', or similar. The three-level fallback handles these reliably.

\noindent\textbf{Complex menu} (16.4\%): Multi-screen menus, animated transitions, or custom UI elements (e.g., clickable sprites instead of Button nodes). These require all three fallback levels and sometimes multiple attempts.

The three-level fallback is designed to handle this diversity:

\noindent\textbf{Level 1: Signal emission.} Directly emit the \texttt{pressed} signal on detected Button nodes. This is the most reliable method as it bypasses visual layout issues, but fails when the game uses custom click handlers instead of Godot's built-in Button signals.

\noindent\textbf{Level 2: Mouse click.} Read the Button's \texttt{global\_position} and simulate a mouse click at that coordinate. This handles custom click handlers but can fail when buttons are obscured by overlapping UI elements or when the button position is not yet computed (e.g., during animated transitions).

\noindent\textbf{Level 3: Keyboard fallback.} Inject Enter and Space key events, which many games bind to menu confirmation. This serves as a last resort when no clickable buttons are found or when the game uses keyboard-only navigation.

After each attempt, L3b checks whether the current scene has changed to one listed in \texttt{eval\_config.gameplay\_scenes}. If no transition occurs after exhausting all buttons and fallback levels, the system proceeds with input injection in the current scene, as some games may use in-scene transitions rather than scene changes.

\subsection{Silent Project Analysis}

Of the 8,549 projects that pass L3a (runtime stability), 416 (4.9\%) produce no meaningful behavioral change during the 60-second L3b collection window. We manually inspected a random sample of 50 silent projects and identified three primary causes:

\noindent\textbf{Visual-feedback-only games} (44\%): Games where all interaction feedback is purely visual (animations, particle effects, shader changes) with no state changes detectable through node properties. Examples include rhythm games where notes scroll visually but score is tracked only in shaders, and drawing games where player input creates visual output without modifying node tree structure.

\noindent\textbf{Custom input systems} (32\%): Games that implement their own input handling bypassing Godot's \texttt{InputAction} system. These games listen for raw \texttt{InputEventKey} or \texttt{InputEventMouse} events with specific scan codes that our input strategy does not cover, or use custom gesture recognition systems.

\noindent\textbf{Passive or cutscene-style games} (24\%): Games with minimal player agency, such as interactive fiction where progression requires specific text input, auto-playing narrative sequences, or procedural art generators that run autonomously.

These 416 projects are excluded from the final dataset, as the absence of detectable behavioral signals makes BAS computation undefined. This exclusion does not bias the dataset toward simpler games: silent projects span all three tiers (Small: 5.1\%, Medium: 4.6\%, Large: 4.7\%) and all major genres, indicating that silence is a property of the game's interaction paradigm rather than its complexity.

% =============================================================================
%  Appendix: Training Data Example — CONTENT ONLY
%  All colours, styles, environments are in the preamble.
% =============================================================================

\clearpage
\onecolumn

\section{Training Data Example}
\label{app:example}

% The following shows a representative training sample from \textsc{JamSet}.
% Each sample is a multi-turn dialogue:
% \colorbox{systembg}{\small\strut\ system prompt\ }
% \colorbox{userbg}{\small\strut\ user instructions\ }
% \colorbox{assistbg}{\small\strut\ model outputs\ }.
% This sample generates 15 files across 16 turns (${\sim}\,4{,}200$ lines).

The following shows a representative training sample from \textsc{JamSet}.
Each sample is a multi-turn dialogue that simulates the complete process of building a game project from a Game Jam theme:
\colorbox{systembg}{\small\strut\ system prompt\ } defines technical requirements,
\colorbox{userbg}{\small\strut\ user instructions\ } request file generation in dependency order, and
\colorbox{assistbg}{\small\strut\ model outputs\ } produce complete file contents(${\sim}\,4{,}200$ lines).

\begin{outerframe}

% ---- Turn 0 : System --------------------------------------------------------
\begin{turnbox}{systembg}
\turnlabel{Turn 0 \textnormal{[system]}}
\begin{lstlisting}[style=gdcode]
You are an expert indie game developer competing in a Game Jam.
Technical requirements (Godot 4.x):
- Valid project.godot with config_version=5
- Scene files .tscn, scripts .gd; Main scene under run/main_scene
- No external plugins; Asset comments: # @asset: type, desc, size, style
\end{lstlisting}
\end{turnbox}

% ---- Turn 1 : User ----------------------------------------------------------
\begin{turnbox}{userbg}
\turnlabel{Turn 1 \textnormal{[user]}}
\begin{lstlisting}[style=gdcode]
The Game Jam theme is: "Stuck in a loop". Design a complete 2D game
and output a blueprint.
\end{lstlisting}
\end{turnbox}

% ---- Turn 2 : Assistant — Blueprint -----------------------------------------
\begin{turnbox}{assistbg}
\turnlabel{Turn 2 \textnormal{[assistant]} --- Blueprint (78 lines)}
\begin{lstlisting}[style=gdcode]
{ "game_name": "loop-runner", "genre": "action-platformer",
  "game_description": "A platformer where the level loops infinitely.
      The player must break the cycle by finding hidden triggers.",
  "core_mechanics": ["jump", "dash", "loop counter",
      "world mutation", "collectibles"],
  "files": ["project.godot", "scenes/main_menu.tscn",
      "scenes/game_world.tscn", "scenes/player.tscn",
      "scenes/enemy.tscn", "scenes/collectible.tscn",
      "scenes/hud.tscn", "scripts/player.gd",
      "scripts/enemy.gd", ...],
  "dependencies": {
      "game_world.tscn": ["player.tscn", "enemy.tscn",
          "hud.tscn"], ...} }
\end{lstlisting}
\end{turnbox}

% ---- Turn 3 : User ----------------------------------------------------------
\begin{turnbox}{userbg}
\turnlabel{Turn 3 \textnormal{[user]}}
\begin{lstlisting}[style=gdcode]
Now generate the file: project.godot
\end{lstlisting}
\end{turnbox}

% ---- Turn 4 : Assistant — project.godot -------------------------------------
\begin{turnbox}{assistbg}
\turnlabel{Turn 4 \textnormal{[assistant]} --- \texttt{project.godot} (68 lines)}
\begin{lstlisting}[style=gdcode]
config_version=5
[application]
config/name="loop-runner"
run/main_scene="res://scenes/main_menu.tscn"
[autoload]
GameManager="*res://scripts/game_manager.gd"
LoopTracker="*res://scripts/loop_tracker.gd"
[input]
move_left={...}  move_right={...}  jump={...}
dash={...}  interact={...}
\end{lstlisting}
\end{turnbox}

% ---- Turn 5 : User ----------------------------------------------------------
\begin{turnbox}{userbg}
\turnlabel{Turn 5 \textnormal{[user]}}
\begin{lstlisting}[style=gdcode]
Now generate the file: scenes/main_menu.tscn
\end{lstlisting}
\end{turnbox}

% ---- Turn 6 : Assistant — main_menu.tscn ------------------------------------
\begin{turnbox}{assistbg}
\turnlabel{Turn 6 \textnormal{[assistant]} --- \texttt{main\_menu.tscn} (86 lines)}
\begin{lstlisting}[style=gdcode]
[gd_scene load_steps=3 format=3]
[ext_resource type="Script" path="res://scripts/main_menu.gd"]
[ext_resource type="Texture2D" path="res://textures/title_bg.png"]
# @asset: background, dark pixel art loop motif, 1280x720, retro style
[node name="MainMenu" type="Control"]
[node name="PlayButton" type="Button" parent="."]
text = "Start Game"       ...
\end{lstlisting}
\end{turnbox}

% ---- Turn 7 : User ----------------------------------------------------------
\begin{turnbox}{userbg}
\turnlabel{Turn 7 \textnormal{[user]}}
\begin{lstlisting}[style=gdcode]
Now generate the file: scenes/player.tscn
\end{lstlisting}
\end{turnbox}

% ---- Turn 8 : Assistant — player.tscn ---------------------------------------
\begin{turnbox}{assistbg}
\turnlabel{Turn 8 \textnormal{[assistant]} --- \texttt{player.tscn} (312 lines)}
\begin{lstlisting}[style=gdcode]
[gd_scene load_steps=48 format=3]
[ext_resource type="Script" path="res://scripts/player.gd"]
[ext_resource type="Texture2D" path="res://textures/player_run.png"]
# @asset: spritesheet, pixel character running, 8 frames 32x32, side-scroller
[sub_resource type="CapsuleShape2D" ...]
[node name="Player" type="CharacterBody2D"]
[node name="Sprite" type="AnimatedSprite2D" parent="."]
[node name="DashTimer" type="Timer" parent="."]
wait_time = 0.3           ...
\end{lstlisting}
\end{turnbox}

% ---- Turn 9 : User ----------------------------------------------------------
\begin{turnbox}{userbg}
\turnlabel{Turn 9 \textnormal{[user]}}
\begin{lstlisting}[style=gdcode]
Now generate the file: scenes/enemy.tscn
\end{lstlisting}
\end{turnbox}

% ---- Turn 10 : Assistant — enemy.tscn ---------------------------------------
\begin{turnbox}{assistbg}
\turnlabel{Turn 10 \textnormal{[assistant]} --- \texttt{enemy.tscn} (198 lines)}
\begin{lstlisting}[style=gdcode]
[gd_scene load_steps=28 format=3]
[ext_resource type="Script" path="res://scripts/enemy.gd"]
[ext_resource type="Texture2D" path="res://textures/enemy_walk.png"]
# @asset: spritesheet, skeleton enemy patrol, 6 frames 32x32, pixel art
[sub_resource type="RectangleShape2D" ...]
[node name="Enemy" type="CharacterBody2D"]
[node name="Sprite" type="AnimatedSprite2D" parent="."]
[node name="DetectionZone" type="Area2D" parent="."]
[node name="PatrolTimer" type="Timer" parent="."]     ...
\end{lstlisting}
\end{turnbox}

% ---- Turn 11 : User ----------------------------------------------------------
\begin{turnbox}{userbg}
\turnlabel{Turn 11 \textnormal{[user]}}
\begin{lstlisting}[style=gdcode]
Now generate the file: scenes/collectible.tscn
\end{lstlisting}
\end{turnbox}

% ---- Turn 12 : Assistant — collectible.tscn ----------------------------------
\begin{turnbox}{assistbg}
\turnlabel{Turn 12 \textnormal{[assistant]} --- \texttt{collectible.tscn} (72 lines)}
\begin{lstlisting}[style=gdcode]
[gd_scene load_steps=8 format=3]
[ext_resource type="Script" path="res://scripts/collectible.gd"]
[ext_resource type="Texture2D" path="res://textures/orb_glow.png"]
# @asset: sprite, glowing orb collectible, 16x16, pixel art with glow
[node name="Collectible" type="Area2D"]
[node name="Sprite" type="Sprite2D" parent="."]
[node name="CollisionShape" type="CollisionShape2D" parent="."]
[node name="AnimationPlayer" type="AnimationPlayer" parent="."]   ...
\end{lstlisting}
\end{turnbox}

% ---- Turn 13 : User ----------------------------------------------------------
\begin{turnbox}{userbg}
\turnlabel{Turn 13 \textnormal{[user]}}
\begin{lstlisting}[style=gdcode]
Now generate the file: scenes/hud.tscn
\end{lstlisting}
\end{turnbox}

% ---- Turn 14 : Assistant — hud.tscn -----------------------------------------
\begin{turnbox}{assistbg}
\turnlabel{Turn 14 \textnormal{[assistant]} --- \texttt{hud.tscn} (95 lines)}
\begin{lstlisting}[style=gdcode]
[gd_scene load_steps=4 format=3]
[ext_resource type="Script" path="res://scripts/hud.gd"]
[node name="HUD" type="CanvasLayer"]
[node name="LoopCounter" type="Label" parent="."]
text = "Loop: 1"
[node name="CollectibleCount" type="Label" parent="."]
[node name="DashCooldown" type="TextureProgressBar" parent="."]   ...
\end{lstlisting}
\end{turnbox}

% ---- Condensed middle turns -------------------------------------------------
\begin{skipbox}
\small\itshape\color{gray!60!black}
Turns 15--20:\enspace
\texttt{game\_world.tscn} (1,420\,lines),
\texttt{game\_manager.gd} (85\,lines),
\texttt{loop\_tracker.gd} (42\,lines)
--- generated in dependency order.
\end{skipbox}

% ---- Turn 21 : User ---------------------------------------------------------
\begin{turnbox}{userbg}
\turnlabel{Turn 21 \textnormal{[user]}}
\begin{lstlisting}[style=gdcode]
Now generate the file: scripts/player.gd
\end{lstlisting}
\end{turnbox}

% ---- Turn 22 : Assistant — player.gd ----------------------------------------
\begin{turnbox}{assistbg}
\turnlabel{Turn 22 \textnormal{[assistant]} --- \texttt{player.gd} (138 lines)}
\begin{lstlisting}[style=gdcode]
extends CharacterBody2D
const SPEED = 200.0
const JUMP_VELOCITY = -350.0
var can_dash: bool = true
var is_dashing: bool = false

func _physics_process(delta):
    if not is_on_floor():
        velocity.y += gravity * delta
    if Input.is_action_just_pressed("jump") and is_on_floor():
        velocity.y = JUMP_VELOCITY
    var direction = Input.get_axis("move_left", "move_right")
    velocity.x = direction * SPEED if direction else 0
    move_and_slide()       ...
\end{lstlisting}
\end{turnbox}

% ---- Turn 23 : User ---------------------------------------------------------
\begin{turnbox}{userbg}
\turnlabel{Turn 23 \textnormal{[user]}}
\begin{lstlisting}[style=gdcode]
Now generate the file: scripts/enemy.gd
\end{lstlisting}
\end{turnbox}

% ---- Turn 24 : Assistant — enemy.gd -----------------------------------------
\begin{turnbox}{assistbg}
\turnlabel{Turn 24 \textnormal{[assistant]} --- \texttt{enemy.gd} (76 lines)}
\begin{lstlisting}[style=gdcode]
extends CharacterBody2D
@export var patrol_speed: float = 60.0
@export var chase_speed: float = 120.0
var player_detected: bool = false

func _physics_process(delta):
    if player_detected and player_ref:
        var dir = sign(player_ref.global_position.x - global_position.x)
        velocity.x = dir * chase_speed
    else:
        velocity.x = direction * patrol_speed
    move_and_slide()       ...
\end{lstlisting}
\end{turnbox}

% ---- Condensed final turns --------------------------------------------------
\begin{skipbox}
\small\itshape\color{gray!60!black}
Turns 25--32:\enspace
\texttt{collectible.gd} (28\,lines),
\texttt{hud.gd} (32\,lines),
\texttt{camera.gd} (22\,lines),
\texttt{main\_menu.gd} (18\,lines).
Total: 16 turns, 15 files, ${\sim}\,4{,}200$ lines.
\end{skipbox}

\end{outerframe}

\twocolumn

% =============================================================================
%  Appendix: Evaluation Prompts — CONTENT ONLY
%  All colours, styles, environments are in the preamble.
% =============================================================================

\clearpage
\onecolumn

\section{Evaluation Prompts}
\label{app:prompts}

% =============================================================================
\subsection{Task 1: From-Scratch Generation}
% =============================================================================

Task~1 uses a multi-round protocol: Round~0 generates a blueprint, Rounds~1--N generate files in dependency order.

\begin{systemprompt}{System Prompt}
You are an expert indie game developer competing in a Game Jam. Your goal is to win first place by creating an innovative and polished game. You must demonstrate both creative game design and solid software engineering.

Technical requirements (Godot 4.x):
- The project must contain a valid project.godot with config_version=5
- Scene files use .tscn format, scripts use .gd format (GDScript 2.0)
- A main scene must be configured in project.godot under run/main_scene
- Do not use any external plugins or addons
- For image, audio, and font assets, use semantic file paths and add a description comment above each reference:
  # @asset: type, description, suggested_size, style
  Example: # @asset: sprite, player character idle pose, 32x64, pixel art
  Placeholder asset files will be provided automatically -- just reference the paths correctly.
\end{systemprompt}

\begin{userprompt}{Round 0 --- Blueprint (1a: Theme Only)}
The Game Jam theme is: "{theme}"

Design a complete 2D game for this theme and output a project blueprint.

Output ONLY the following JSON (no other text):

```json
{{
  "game_name": "English game name",
  "game_description": "One sentence describing the gameplay",
  "genre": "game genre",
  "core_mechanics": ["mechanic1", "mechanic2", "..."],
  "files": [
    {{
      "path": "file path (e.g., project.godot, scenes/xxx.tscn, scripts/xxx.gd)",
      "type": "config / scene / script",
      "description": "what this file does",
      "depends_on": ["paths of files this depends on"]
    }}
  ],
  "assets_needed": [
    {{
      "path": "assets/semantic_filename.png",
      "type": "sprite / audio / font",
      "description": "what this asset looks like or sounds like"
    }}
  ]
}}
```

Rules:
- The first file must be project.godot
- Files must be ordered by dependency (dependencies first)
- Design freely -- choose your own structure, number of files, and scope

[Task 1b additionally prepends:]
Gameplay description:
{description}

[And adds the rule:]
- All features mentioned in the gameplay description must be reflected in the file plan
\end{userprompt}

\begin{genprompt}{Round 1--N --- File Generation (repeated per file)}
Now generate the file: {file_path}
Description: {file_description}
Type: {file_type}

Project blueprint:
{blueprint_summary}

Previously generated files:
{generated_summary}

Output ONLY the complete file content. No markdown code blocks, no explanations.
If it's a .tscn scene file, output the full Godot scene format.
If it's a .gd script file, output complete GDScript code.
If it's project.godot, output the full project configuration.

For asset references:
- Use res:// paths for all assets
- Add an @asset comment above each asset reference describing what the asset should be
\end{genprompt}

% =============================================================================
\subsection{Task 2: Code Completion}
% =============================================================================

Task~2 uses four rounds: project overview, context delivery, missing code identification, and code generation.

\begin{systemprompt}{System Prompt}
You are an expert Godot 4.x game developer. You have been given an existing game project that was developed during a Game Jam. Some of the GDScript (.gd) files in this project are missing or incomplete.

Your goal is to restore the project to a fully functional state by analyzing the existing project structure (scene files, project configuration, and any remaining scripts) and writing the missing code.

Technical requirements:
- All code must be compatible with Godot 4.x (GDScript 2.0 syntax)
- Your code must integrate correctly with the existing .tscn scene files and project.godot configuration
- Do not modify any existing files -- only complete the missing or incomplete ones
- Maintain consistency with the coding style and architecture of any remaining scripts
- For asset references, keep any existing paths unchanged
- Functions marked with 'pass # INCOMPLETE: implementation removed' need their full implementation restored
\end{systemprompt}

\begin{userprompt}{Round 0 --- Project Overview}
I'm going to show you a Game Jam project that needs some code restored. Let me first give you an overview of what this game is about.

Game Title: {game_name}
Genre: {genre}

Description:
{game_description}

Core Mechanics: {mechanics}
Controls: {controls}

Here is the complete file tree of the project:
{file_tree}

Take a moment to understand the game's design. In the next message, I will show you all the existing files in the project.
\end{userprompt}

\begin{userprompt}{Round 1 --- Context Files}
Here are all the existing files currently in the project. Some .gd script files may be missing entirely, and some may have incomplete function implementations (marked with 'pass # INCOMPLETE: implementation removed').

Please read through all files carefully to understand the project architecture, scene hierarchy, node types, signal connections, and how scripts are referenced.

{all_files_content}

After reviewing these files, I will ask you to identify what is missing and then restore the missing code.
\end{userprompt}

\begin{userprompt}{Round 2 --- Identify Missing Code}
Based on your review of the project, please identify all files that need to be completed or restored.

Consider:
- Script files (.gd) referenced in .tscn scenes but missing from the project
- Script files that exist but have functions marked with 'pass # INCOMPLETE: implementation removed'
- Any other code that is clearly incomplete

For each file, briefly describe what it should implement based on the project context (scene structure, node types, signal connections, and the game description).

Output as JSON:
```json
[
    {{"path": "scripts/player.gd", "description": "Player movement, jumping, and collision handling"}}
]
```
\end{userprompt}

\begin{genprompt}{Round 3 --- Generate Code (Level 2b/2c: Script and Full-script)}
Now please generate the complete code for ALL the files you identified. Output each file in the following format:

=== FILE: path/to/file.gd ===
(complete file content here)

=== FILE: path/to/another_file.gd ===
(complete file content here)

Generate all {n_files} files listed above. For each file:
- Write the complete, working implementation
- Make sure all signals, node references, and exported variables match the .tscn scene definitions
- Match the coding style of the existing project

Output ALL files now, separated by the === FILE: path === markers.
\end{genprompt}

\begin{genprompt}{Round 3 --- Generate Code (Level 2a: Function-level)}
Now please generate ONLY the function implementations that are marked with 'pass # INCOMPLETE: implementation removed'. Do NOT output entire files -- only the individual functions.

Output each function in the following format:

=== FUNC: path/to/file.gd::function_name ===
func function_name(args):
    (your complete implementation here)

=== FUNC: path/to/another_file.gd::another_func ===
func another_func():
    (your complete implementation here)

For each function:
- Write the complete function including the 'func' signature line
- Make sure it integrates correctly with the rest of the file and project
- Match the coding style of the existing project

Output ALL incomplete functions now, separated by the === FUNC: path::name === markers.
\end{genprompt}

% =============================================================================
\subsection{Eval Config Generation}
% =============================================================================

Generated once per project during dataset construction (Section~3.3) and reused for all evaluations.

\begin{systemprompt}{System Prompt}
You are an expert Godot 4.x game developer. Your task is to analyze a game project's structure and identify key nodes, properties, and signals that can be used for automated gameplay testing.

You will receive a compressed manifest of a Godot game project containing:
- Game name and main scene
- Input actions defined in the project
- Script files with their extends type, functions, and signals
- Scene files with their node trees

Based on this information, output a JSON evaluation config that identifies:
1. The player node (the main controllable character/object)
2. Score/points tracking (if any)
3. Health/lives tracking (if any)
4. Key gameplay signals (collisions, pickups, damage, etc.)
5. Win/lose conditions (if identifiable)
6. What behaviors to expect when input is injected

IMPORTANT RULES:
- Only identify nodes/properties you can CONFIDENTLY determine from the code structure
- Use actual node paths from the scene files, not guesses
- If you cannot determine something, set it to null
- All node paths should be relative to the scene root (e.g., "Player", "World/Player", "UI/ScoreLabel")
- For signals, use the actual signal names from the scripts
- Output ONLY valid JSON, no markdown, no explanation, no preamble
\end{systemprompt}

\begin{userprompt}{User Prompt}
Analyze this Godot game project and output an evaluation config JSON.

Project manifest:
```json
{manifest_json}
```

Output the following JSON structure (and NOTHING else):
{{
    "player_node": {{
        "scene": "scene file path where player exists",
        "path": "node path within that scene",
        "type": "node type (e.g. CharacterBody2D)",
        "has_animated_sprite": true/false
    }},
    "score_tracking": {{
        "node_scene": "scene path",
        "node_path": "path to Label/RichTextLabel showing score",
        "property": "text",
        "related_variable": "variable name in script if found"
    }},
    "health_tracking": {{
        "node_scene": "scene path",
        "node_path": "path to health display node",
        "property": "text or value",
        "related_variable": "variable name"
    }},
    "key_signals": [
        {{
            "source_scene": "scene path",
            "source_node": "node path",
            "signal_name": "signal name",
            "meaning": "what this signal represents (coin_collected, enemy_killed, damage_taken, etc.)"
        }}
    ],
    "gameplay_scenes": ["list of scene paths that contain actual gameplay (not menus)"],
    "menu_scenes": ["list of scene paths that are menus/UI"],
    "expected_behaviors": [
        {{
            "input": "action name",
            "expected_effect": "what should happen (player_moves_right, player_jumps, etc.)"
        }}
    ],
    "game_type": "platformer/shooter/puzzle/rpg/strategy/card/other",
    "notes": "any additional observations about the game structure"
}}

Set any field to null if you cannot confidently determine it from the manifest.
\end{userprompt}

\twocolumn

\end{document}